\documentclass[fleqn,usenatbib]{mnras}

\usepackage{newtxtext,newtxmath}

\usepackage[T1]{fontenc}
\usepackage[utf8]{inputenc}
\usepackage{ae,aecompl}
\usepackage{enumitem}

\usepackage{graphicx}	
\usepackage[singlelinecheck=off]{subcaption}
\usepackage{amsmath}	
\usepackage{amssymb}	
\usepackage{adjustbox}
\usepackage{nccmath}
\usepackage{color}
\usepackage[displaymath]{lineno}

\DeclareMathSizes{12}{12}{10}{8}
\title[Rapidly evolving transients in DES]{Rapidly evolving transients in the Dark Energy Survey}

\author[DES Collaboration]{
\parbox{\textwidth}{
\Large
M.~Pursiainen$^{1}$,
M.~Childress$^{1}$,
M.~Smith$^{1}$,
S.~Prajs $^{1}$,
M.~Sullivan$^{1}$,
T.~M.~Davis$^{2,3}$,
R.~J.~Foley$^{4}$,
J.~Asorey$^{2,5,3}$,
J.~Calcino$^{3}$,
D.~Carollo$^{2}$,
C.~Curtin$^{5}$,
C.~B.~D'Andrea$^{6}$,
K.~Glazebrook$^{5}$,
C.~Gutierrez$^{1}$,
S.~R.~Hinton$^{3}$,
J.~K.~Hoormann$^{3}$,
C.~Inserra$^{1}$,
R.~Kessler$^{7}$,
A.~King$^{3}$,
K.~Kuehn$^{8}$,
G.~F.~Lewis$^{9}$,
C.~Lidman$^{8,10}$,
E.~Macaulay$^{3}$,
A.~M\"oller$^{2,10}$,
R.~C.~Nichol$^{11}$,
M.~Sako$^{6}$,
N.~E.~Sommer$^{2,10}$,
E.~Swann$^{11}$,
B.~E.~Tucker$^{2,10}$,
S.~A.~Uddin$^{2,12}$,
P.~Wiseman$^{1}$,
B.~Zhang$^{2,10}$,
T.~M.~C.~Abbott$^{13}$,
F.~B.~Abdalla$^{14,15}$,
S.~Allam$^{16}$,
J.~Annis$^{16}$,
S.~Avila$^{11}$,
D.~Brooks$^{14}$,
E.~Buckley-Geer$^{16}$,
D.~L.~Burke$^{17,18}$,
A.~Carnero~Rosell$^{19,20}$,
M.~Carrasco~Kind$^{21,22}$,
J.~Carretero$^{23}$,
F.~J.~Castander$^{24,25}$,
C.~E.~Cunha$^{17}$,
C.~Davis$^{17}$,
J.~De~Vicente$^{26}$,
H.~T.~Diehl$^{16}$,
P.~Doel$^{14}$,
T.~F.~Eifler$^{27,28}$,
B.~Flaugher$^{16}$,
P.~Fosalba$^{24,25}$,
J.~Frieman$^{16,7}$,
J.~Garc\'ia-Bellido$^{29}$,
D.~Gruen$^{17,18}$,
R.~A.~Gruendl$^{21,22}$,
G.~Gutierrez$^{16}$,
W.~G.~Hartley$^{14,30}$,
D.~L.~Hollowood$^{4}$,
K.~Honscheid$^{31,32}$,
D.~J.~James$^{33}$,
T.~Jeltema$^{34}$,
N.~Kuropatkin$^{16}$,
T.~S.~Li$^{16,7}$,
M.~Lima$^{35,19}$,
M.~A.~G.~Maia$^{19,20}$,
P.~Martini$^{31,36}$,
F.~Menanteau$^{21,22}$,
R.~L.~C.~Ogando$^{19,20}$,
A.~A.~Plazas$^{28}$,
A.~Roodman$^{17,18}$,
E.~Sanchez$^{26}$,
V.~Scarpine$^{16}$,
R.~Schindler$^{18}$,
R.~C.~Smith$^{13}$,
M.~Soares-Santos$^{37}$,
F.~Sobreira$^{38,19}$,
E.~Suchyta$^{39}$,
M.~E.~C.~Swanson$^{22}$,
G.~Tarle$^{40}$,
D.~L.~Tucker$^{16}$,
A.~R.~Walker$^{13}$
\begin{center} (DES Collaboration) \end{center}
}
\vspace{0.4cm}
\\
\parbox{\textwidth}{
$^{1}$ School of Physics and Astronomy, University of Southampton,  Southampton, SO17 1BJ, UK\\
$^{2}$ ARC Centre of Excellence for All-sky Astrophysics (CAASTRO)\\
$^{3}$ School of Mathematics and Physics, University of Queensland,  Brisbane, QLD 4072, Australia\\
$^{4}$ Department of Astronomy and Astrophysics, University of California, Santa Cruz, CA 95064, USA\\
$^{5}$ Centre for Astrophysics \& Supercomputing, Swinburne University of Technology, Victoria 3122, Australia\\
$^{6}$ Department of Physics and Astronomy, University of Pennsylvania, Philadelphia, PA 19104, USA\\
$^{7}$ Kavli Institute for Cosmological Physics, University of Chicago, Chicago, IL 60637, USA\\
$^{8}$ Australian Astronomical Observatory, North Ryde, NSW 2113, Australia\\
$^{9}$ Sydney Institute for Astronomy, School of Physics, A28, The University of Sydney, NSW 2006, Australia\\
$^{10}$ The Research School of Astronomy and Astrophysics, Australian National University, ACT 2601, Australia\\
$^{11}$ Institute of Cosmology \& Gravitation, University of Portsmouth, Portsmouth, PO1 3FX, UK\\
$^{12}$ Purple Mountain Observatory, Chinese Academy of Sciences, Nanjing, Jiangshu 210008, China\\
$^{13}$ Cerro Tololo Inter-American Observatory, National Optical Astronomy Observatory, Casilla 603, La Serena, Chile\\
$^{14}$ Department of Physics \& Astronomy, University College London, Gower Street, London, WC1E 6BT, UK\\
$^{15}$ Department of Physics and Electronics, Rhodes University, PO Box 94, Grahamstown, 6140, South Africa\\
$^{16}$ Fermi National Accelerator Laboratory, P. O. Box 500, Batavia, IL 60510, USA\\
$^{17}$ Kavli Institute for Particle Astrophysics \& Cosmology, P. O. Box 2450, Stanford University, Stanford, CA 94305, USA\\
$^{18}$ SLAC National Accelerator Laboratory, Menlo Park, CA 94025, USA\\
$^{19}$ Laborat\'orio Interinstitucional de e-Astronomia - LIneA, Rua Gal. Jos\'e Cristino 77, Rio de Janeiro, RJ - 20921-400, Brazil\\
$^{20}$ Observat\'orio Nacional, Rua Gal. Jos\'e Cristino 77, Rio de Janeiro, RJ - 20921-400, Brazil\\
$^{21}$ Department of Astronomy, University of Illinois at Urbana-Champaign, 1002 W. Green Street, Urbana, IL 61801, USA\\
$^{22}$ National Center for Supercomputing Applications, 1205 West Clark St., Urbana, IL 61801, USA\\
$^{23}$ Institut de F\'{\i}sica d'Altes Energies (IFAE), The Barcelona Institute of Science and Technology, Campus UAB, 08193 Bellaterra (Barcelona) Spain\\
$^{24}$ Institut d'Estudis Espacials de Catalunya (IEEC), 08193 Barcelona, Spain\\
$^{25}$ Institute of Space Sciences (ICE, CSIC),  Campus UAB, Carrer de Can Magrans, s/n,  08193 Barcelona, Spain\\
$^{26}$ Centro de Investigaciones Energ\'eticas, Medioambientales y Tecnol\'ogicas (CIEMAT), Madrid, Spain\\
$^{27}$ Department of Astronomy/Steward Observatory, 933 North Cherry Avenue, Tucson, AZ 85721-0065, USA\\
$^{28}$ Jet Propulsion Laboratory, California Institute of Technology, 4800 Oak Grove Dr., Pasadena, CA 91109, USA\\
$^{29}$ Instituto de Fisica Teorica UAM/CSIC, Universidad Autonoma de Madrid, 28049 Madrid, Spain\\
$^{30}$ Department of Physics, ETH Zurich, Wolfgang-Pauli-Strasse 16, CH-8093 Zurich, Switzerland\\
$^{31}$ Center for Cosmology and Astro-Particle Physics, The Ohio State University, Columbus, OH 43210, USA\\
$^{32}$ Department of Physics, The Ohio State University, Columbus, OH 43210, USA\\
$^{33}$ Harvard-Smithsonian Center for Astrophysics, Cambridge, MA 02138, USA\\
$^{34}$ Department of Astronomy and Astrophysics, University of California, Santa Cruz, CA 95064, USA\\
$^{35}$ Departamento de F\'isica Matem\'atica, Instituto de F\'isica, Universidade de S\~ao Paulo, CP 66318, S\~ao Paulo, SP, 05314-970, Brazil\\
$^{36}$ Department of Astronomy, The Ohio State University, Columbus, OH 43210, USA\\
$^{37}$ Brandeis University, Physics Department, 415 South Street, Waltham MA 02453\\
$^{38}$ Instituto de F\'isica Gleb Wataghin, Universidade Estadual de Campinas, 13083-859, Campinas, SP, Brazil\\
$^{39}$ Computer Science and Mathematics Division, Oak Ridge National Laboratory, Oak Ridge, TN 37831\\
$^{40}$ Department of Physics, University of Michigan, Ann Arbor, MI 48109, USA\\
}
}

\date{Accepted XXX. Received YYY; in original form ZZZ}

\pubyear{2018}

\begin{document}
\label{firstpage}
\pagerange{\pageref{firstpage}--\pageref{lastpage}}
\maketitle

\begin{abstract}
We present the results of a search for rapidly evolving transients in the Dark Energy Survey Supernova Programme. These events are characterized by fast light curve evolution (rise to peak in $\lesssim 10$ d and exponential decline in $\lesssim30$ d after peak). We discovered 72 events, including 37 transients with a spectroscopic redshift from host galaxy spectral features. The 37 events increase the total number of rapid optical transients by more than factor of two. They are found at a wide range of redshifts ($0.05<z<1.56$) and peak brightnesses ($-15.75>M_\mathrm{g}>-22.25$). The multiband photometry is well fit by a blackbody up to few weeks after peak. The events appear to be hot ($T\approx10000-30000$ K) and large ($R\approx 10^{14}-2\cdot10^{15}$ cm) at peak, and generally expand and cool in time, though some events show evidence for a receding photosphere with roughly constant temperature. Spectra taken around peak are dominated by a blue featureless continuum consistent with hot, optically thick ejecta. We compare our events with a previously suggested physical scenario involving shock breakout in an optically thick wind surrounding a core-collapse supernova (CCSNe), we conclude that current models for such a scenario might need an additional power source to describe the exponential decline. We find these transients tend to favor star-forming host galaxies, which could be consistent with a core-collapse origin.  However, more detailed modeling of the light curves is necessary to determine their physical origin.

\end{abstract}




\section{Introduction}

Dedicated wide field supernova (SN) surveys are discovering large numbers of traditional types of SNe, including both Type Ia produced in thermonuclear disruptions of white dwarfs and Type II/Ibc originating in the aftermath of core-collapse of massive stars ($\gtrsim8M_{\odot}$) (see, e.g., \citealt{Filippenko1997} or \citealt{Gal-Yam2017}, for review). However, as the coverage in depth, area and time has improved, the surveys have also started to detected different types of exotic optical transients such as superluminous SNe (SLSNe)(e.g., \citealt{Quimby2011}; see \citealt{Howell2017}, for review) and Ca-rich transients \citep[e.g.][]{Perets2010}, with behavior explained by different physical mechanisms than those used for typical SNe. 

Over the last decade a new interesting class of transient with rapid photometric evolution has been discovered. These events can be luminous ($-15>M>-20$) and they are characterized by fast light curve evolution with rise to maximum brightness in $\lesssim10$ d and exponential decline in $\lesssim30$ d after the peak in rest frame, making them very difficult to observe. Searches are often designed to discover and follow Type Ia SNe due to their use as cosmological probes, for which high cadence is not necessarily required. Due to this, the rapid events can be detected but it is difficult to characterize them early enough for follow-up observations.  To complicate the task even more, these events appear to be very rare, with a (magnitude-limited) rate of only 4\%-7\% of that of the core-collapse SNe (CCSNe) as estimated by \citet{Drout2014} based on a sample of 10 found in Pan-STARRS1. 

The net result of these factors is that only few tens of rapidly evolving transients have been discovered to date.  The majority of these come from modern wide-field surveys, including a sample of fast blue transients discovered by Pan-STARRS1 \citep{Drout2014}, some rapidly rising luminous events discovered by the Supernova Legacy Survey \citep{Arcavi2016}, a few rapidly rising events discovered by Subaru Hyper Suprime-Cam Transient Survey for which only the rise is observed \citep{Tanaka2016}, a few individual events from the Palomar Transient Factory such as PTF 09uj \citep{Ofek2010} and iPTF 16asu \citep{Whitesides2017}, and a single candidate from Subaru HIgh-Z sUpernova CAmpaign \citep[SHIZUCA, HSC17dbpf,][]{Curtin2018, Moriya2018}. Some additional events with classical spectroscopic classifications have also shown rapid evolution in their light curves:  Type Ibn SN 1999cq \citep{Matheson2000a}, Type Ib SN2002bj \citep{Poznanski2009}, Type Ic SN2005ek \citep{Drout2013},  Type Ibn SN2015U \citep{Pastorello2015a, Tsvetkov2015, Shivvers2016} and Type Ibn SN iPTF 15ul \citep{Hosseinzadeh2017}. The Kepler space telescope also discovered a similar transient KSN 2015K, with observing cadence of 30 minutes \citep{Rest2018}. Two additional noteworthy events are ``Dougie’’ \citep{Vinko2015} and SN2011kl \citep{Greiner2015}. ``Dougie’’ was an exceptionally bright ($M_\mathrm{R}\approx-23$) rapidly rising event, while SN2011kl was fainter and slightly slower, but it was identified to be an afterglow of ultra-long gamma-ray burst \citep[GRB, ][]{Gendre2013, Stratta2013, Levan2014}.  The recently discovered neutron star merger GW170817 was also found to have a rapidly fading optical counterpart \citep[see][and references therein]{LIGOScientificCollaboration2017}, though this was significantly faster than the Pan-STARRS sample of rapid transients presented by \citet{Drout2014}. It is worth mentioning that the broad category of rapidly evolving transients appears to be heterogeneous, as illustrated by the diversity of observed spectroscopic features: for example, iPTF 16asu displayed a Ic-BL like spectrum \citep{Whitesides2017} while PTF 10iam showed hydrogen in its spectrum \citep{Arcavi2016}.

The majority of these events appear to have similar light curve evolution, but have a wide range of luminosities that are difficult to describe with any individual model. For instance the sample of \citet{Drout2014} is described by time above half maximum $t_\mathrm{1/2}\lesssim12$ d despite having peak brightnesses in range $-17>M_\mathrm{g}>-20$. Several different scenarios have been considered to explain the rapid light curve evolution including the explosion of a stripped massive star \citep{Drout2013,Tauris2013,Kleiser2014} and shock breakout in either a dense circumstellar medium \citep[CSM,][]{Ofek2010} or an extended low-mass stellar envelope \citep{Drout2014}, white dwarf detonation \citep{Arcavi2016}, CSM interaction \citep{Arcavi2016} and magnetar spin-down \citep{Arcavi2016}. While some models have been disfavored after comparison to data \citep[e.g.  magnetar spin-down,][]{Arcavi2016}, a number of models could still viably explain the observed properties of these rapid transients. Characterizing the distribution of the observed properties of rapid transients remains limited by the low number of discovered objects, which in turn limits the ability to constrain their physical origin.

In this work we present a new large sample (72) of rapidly evolving transients discovered by the Dark Energy Survey Supernova Program (DES-SN) during its first four years of operation. DES-SN has a cadence of roughly one week and therefore only up to 4-5 measurements of individual transients can be expected. However, as DES-SN provides deep ($m_\mathrm{lim} \sim 24$~mag) multi-band ($griz$) photometry probing a large volume, finding a significant number of such events is possible.  We discovered in total 72 events in DES-SN, including 37 events that have spectroscopic redshifts from host galaxy spectral features present in a spectrum of the host or the transient itself. This spectroscopic sample increases the total number of rapid transients ($t_\mathrm{1/2}\lesssim 12$ d) by more than factor of two. The sample has a wide luminosity range from $M_\mathrm{g}\approx-15.75$ to $M_\mathrm{g}\approx-22.25$.

This paper is the first investigation of rapidly evolving events in DES-SN, and as such we focus here primarily on the discovery and simple analysis of these transients in our survey. In this work, our transient discovery technique is built on a simple characterization of all unclassified short lived (less than few months) DES-SN transients using linear and Gaussian fits to the first four years of DES-SN photometric data. We use linear fits on data outside the event to reject events exhibiting coherent long term variability, which likely are active galactic nuclei (AGN).  We employ Gaussian fits to the DES-SN transient events themselves to estimate crude duration of the event in order to isolate short-duration transients. A follow-up analysis is underway and will be presented in a second paper, where we will include the full five-year sample of rapidly evolving transients in DES-SN, perform rigorous tests of our selection techniques to reduce amount of visual selection in the analysis,  estimate the rate of these events, and attempt to describe their light curve behavior in the context of physical models such as shock-breakout models. 

The paper is structured as follows. In Section \ref{sec:obs} we give a brief overview of DES and a description of the sample selection, and summarize the observations used in our analysis. In Section \ref{sec:sample} we describe our sample and then in Sections \ref{sec:photo}-\ref{sec:spectra} we present an analysis of their photometric and spectroscopic properties. The host galaxy properties are presented in Section \ref{sec:hostgal} and finally in Section \ref{sec:discussions} we discuss the differences between our events and the previously proposed shock breakout scenario. In Section \ref{sec:conclusions} we conclude our paper.  Throughout this paper we calculate distances assuming a flat $\Lambda$CDM cosmology with $\Omega_\mathrm{M}=0.3$ and H$_\mathrm{0}$ = 70 km s$^{-1}$ Mpc$^{-1}$.

\section{Observations and Sample Selection}

\begin{figure*}
\includegraphics[width=0.98\textwidth]{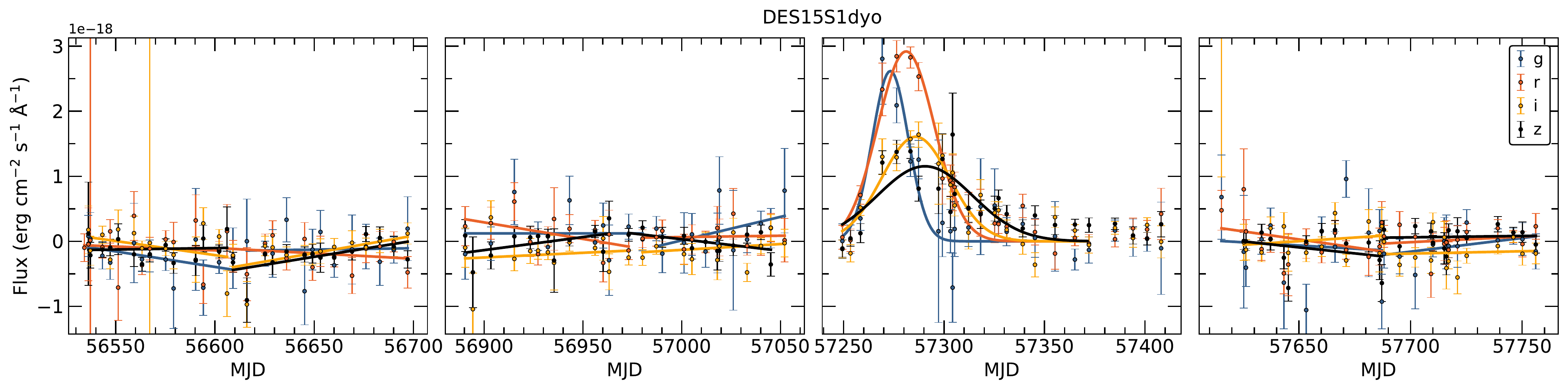}
\caption{4-year light curve of DES15S1dyo, a likely Type Ia SN, showing the linear and Gaussian fits. Linear fits are used to characterize the background variability outside the event year and Gaussian fits the rough duration of the event itself.}
\label{fig:pipeline}
\end{figure*}

\label{sec:obs}
\subsection{Dark Energy Survey Overview}
The Dark Energy Survey \citep[DES, ][]{Flaugher2005} is an international collaboration with a primary aim to study cosmic acceleration, and thereby constrain the nature of the mysterious dark energy which drives it. DES constructed the highly sensitive Dark Energy Camera \citep[DECam,][]{Flaugher2015}, which has a 3 deg$^2$ field of view and is mounted on the 4 m Victor M. Blanco telescope at Cerro Tololo Inter-American Observatory (CTIO) in Chile. For providing the camera, the DES collaboration was awarded 105 nights a year for five years, starting from August 2013. DES recently concluded its fifth observing year in February 2018.

The DES main survey observes a wide area of about 5000 deg$^2$ (roughly 1/8 of the sky), 
and at the conclusion of its fourth year of observations DES has visited each 3 deg$^2$ field within this footprint at least six times. Additionally, DES-SN searches for transients in ten 3 deg$^2$ fields (i.e. ten single pointings of DECam). These ten fields are observed in \textit{griz} bands roughly once a week, with the goal of discovering and photometrically following Type Ia SNe for cosmology \citet{Bernstein2012}. Eight of these fields are referred to as “shallow” as they are observed with exposure times of only few minutes, while the remaining two are called “deep” as they are observed three to nine times longer depending on the band. The events in deep fields can be recognized by “C3” or “X3” in the event name (e.g DES15C3opk). In the shallow fields all four bands are observed nearly simultaneously, but in the deep fields there may be several days between the observations in different bands. For more information on the DES-SN observing strategy see D'Andrea et al. (2018, in prep.), and on the DES-SN difference-imaging search pipeline and transient identification algorithms see \citet{Kessler2015} and \citet{Goldstein2015}.

Out of the multi-year light curves of DES-SN transients, we use the Science Verification (SV) data as a reference for transients detected within the first year of the survey. Then we use the first year data as reference for the second year, and finally all subsequent transients detected within years three to five use data from the second season as a reference. Thus any transient in the reference data would show up as a constant offset to the photometry, making the subtraction fluxes too low. Thus in theory we may have missed any transient which is coincident with a transient that occurred in the reference epoch data, but this should not affect the shape of the light curves we measure for the sample analyzed in this paper.

\subsection{Automated Classification of SN-like events}

In its first four years DES-SN has detected roughly 16000 objects classified as single-season events (i.e. objects detected in image subtractions during one year but not the others). These objects include different varieties of SNe, but also significant amounts of AGN and some spurious detections. In order to find peculiar SNe, such as the rapidly evolving ones, it is important to cut down the number of viable candidates automatically before visual inspection of the light curves.  Therefore, we devised a photometric classification pipeline, \textit{ClassPipe}, based on linear and Gaussian fits on the light curves of four bands ($griz$) to quantify the light curve shape.  All fitting in this paper has been performed with \texttt{LMFIT} package for python \citep{Newville2014}.

First, \textit{ClassPipe} looks for SNe by requiring that the light curve during the three seasons outside of the event year (the year when event was formally discovered and named) should be non-variable. We split each of three season in half ($\sim75$d each) and found the average slope of the linear fits over the four bands for each of the six intervals. In the second step we fitted the light curves of each band of the event season with Gaussian profiles to characterize the shape of the event. The parameters of Gaussian fits used to classify the events were the amplitude $A$, MJD of the peak, and the Full Width at Half Maximum (FWHM). The cut-off values for parameters of the linear and Gaussian fits were defined empirically based on the spectroscopically confirmed SNe and AGN from first three years of DES-SN (see Table \ref{tab:pipeline_test}).  

We note here that the choice of a Gaussian fit to characterize the transient light curve width may bias our results away from transient with rapidly rising but highly asymmetric light curves.  A broader range of light curve parameterizations will be explored in a future analysis.  We do note, however, that this simplistic Gaussian parametrization resulted in a high success rate of detecting spectroscopically-confirmed transients (see discussion below and Table \ref{tab:pipeline_test}).

An object classified as ``SN-like'' had to fulfill the following conditions:

\begin{enumerate}[labelsep=0.2cm, leftmargin=*]
\item The second highest absolute value of the six mean slopes of the linear fits outside the event year was smaller than $10^{-20}~\mathrm{erg}~\mathrm{cm}^{-2}~\mathrm{s}^{-1}~\mathrm{d}^{-1}$~\AA$^{-1}$ (the highest value was ignored as a means of outlier rejection). If there was not enough data to achieve at least two out of six mean slopes, the object was rejected. In such cases either subtracting the host galaxy emission failed and corresponding images were not used or the event was found near the edge of a chip and hence image might not always contain the transient due to small dithering in the pointings. 
\item The peaks of Gaussian fits on the event year for at least three bands were within 15d from the mean MJD of the peaks.
\item  The peak fluxes of Gaussian fits on the event year were higher than  $10^{-19} ~\mathrm{erg}~\mathrm{cm}^{-2}~\mathrm{s}^{-1}$~\AA$^{-1}$  in at least three bands, or or higher than $5\cdot10^{-19}~\mathrm{ erg}~\mathrm{cm}^{-2}~\mathrm{s}^{-1}$~\AA$^{-1}$  in at least one band. If the latter was true for only one band, the FWHM of that band also had to be between 12 to 95 days.
\item The mean observed flux of the event year over four bands was higher than the mean observed flux of any other season.
\end{enumerate}

The limiting values of flux in Cut 3. correspond to magnitudes of $m_\mathrm{griz}=26.7$, $26.0$, $25.6$, $25.3$ when the peak of Gaussian fit  is brighter in at least three bands and $m_\mathrm{griz}=25.0$, $24.3$, $23.9$, $23.5$ in case of only one band. The typical average depths in DES-SN, measured as the magnitude where 50\% of inserted fakes are detected, are $m_\mathrm{griz}=24.3$, $24.6$, $24.5$, $24.3$ in the deep fields and $m_\mathrm{griz}=23.6$, $23.4$, $23.2$, $23.1$ in the shallow fields \citep{Kessler2015}. Our cutoff limits are considerably lower as we do not want to lose any interesting events, regardless how faint they are.

 \textit{ClassPipe} also looks for events that were either at the beginning or at the end of the event year, where only the decline or the rise was observed. This was done because missing the rise would result in much wider Gaussian fits and missing the tail in much shorter. Therefore a real SN could end up being classified as non-SN. Through visual inspection of light curves, we found the following criteria helped classify an event caught on the ``decline'':

\begin{itemize}[labelsep=0.2cm, leftmargin=*]
\item In at least two bands, the peak flux was the first data point of the year, and the second highest flux was one of the first three data points, or
\item the peak flux was the first data point in one band and within first three data points in the other three bands, and the second highest fluxes were one of the first three data points in each band, or
\item the peak MJD of a Gaussian fit was before the first data point of the observing season in at least two bands.   
\end{itemize}

As the rise of a SN light curve is typically much shorter in duration than the decline, our requirements for it were much stricter.  We classified an event as caught on the ``rise'' if: 
\begin{itemize}[labelsep=0.2cm, leftmargin=*]
\item In at least three bands, the peak flux of a band was the last data point of the year, and the second highest flux was within the last three data points, or
\item the peak MJD of a Gaussian fit was after the last data point of the event year in at least two bands.   
\end{itemize}

To demonstrate how  \textit{ClassPipe} works we have plotted a 4-year light curve of DES15S1dyo, a likely Type Ia SN that passed our set of criteria, in Figure \ref{fig:pipeline}. We tested the performance of  \textit{ClassPipe} on spectroscopically confirmed SNe and AGN from first three years of DES-SN with several combinations of parameter values to improve its performance before settling on the final values of cuts presented above. This also included a test on how well \textit{ClassPipe} differentiated SNe as SN-like. For this we defined ``SN range" $12 ~\mathrm{d} \leq \mathrm{FWHM} \leq 95 ~\mathrm{d}$, where either the mean FWHM of the Gaussian fits or the individual FWHMs of three bands had to be for the event to be classified as ``SNe" For all cases which did not satisfy one of the three sets of criteria above (``decline", ``rise", or ``SN"), the object was classified as ``non-SNe". Table \ref{tab:pipeline_test} summarizes the performance of  \textit{ClassPipe} with the final cut values on this spectroscopically confirmed sample of objects. From these results we can conclude that  \textit{ClassPipe} finds Type Ia SNe with high completeness with only 9/251 classified as non-SNe. These nine SNe have strong variation in the host galaxy emission and hence they were rejected during the linear fits phase. However, the FWHM is not a good parameter to identify longer-lasting SNe such as type II from AGN as such events have often comparable FWHMs. This leads to a high rate of misclassification for Type II SNe (7/34 classified as non-SNe) even if AGN are excluded as well with only 1/50 AGN classified as SNe.

In Table \ref{tab:pipeline} we have presented the results of  \textit{ClassPipe} for the full four-year sample of unclassified DES-SN transients that occurred only during one observing season. The table gives the number of events cut at different phases of the  \textit{ClassPipe}. Altogether roughly half of the of the events passed  \textit{ClassPipe}, now having a crude measurement of the duration of the event in the form of FWHM.

\begin{table}
\centering
\def\arraystretch{1.1}
\normalsize 
\caption{Classifications obtained with \textit{ClassPipe} on the spectroscopically classified SNe (Ia, II and Ibc) and AGN from first three years of DES-SN. The pipeline classifies the events as ``SN-like", not ``SN-like" or events that were observed only on the decline or the rise, respectively. For more details see text.}
\label{tab:pipeline_test}
	\begin{tabular}{l | l l l l}
		\hline
		Type		&	SN Ia &	SN II	&	SN Ibc	&	AGN	\\
		\hline
		Total		&	251	&	34	&	9	&	49 \\
		\hline		
		Non-SNe	&	9	&	7	&	1	&	47	\\
		SNe		&	232	&	20	&	6	&	1	\\
		Decline	&	10	&	6	&	2	&	1	\\
		Rise		&	0	&	1	&	0	&	0	\\
		\hline		
\end{tabular}
\end{table}

\begin{table}
\centering
\def\arraystretch{1.1}
\small
\caption{Results of the classification pipeline, \textit{ClassPipe}, on four years of DES-SN data showing the number of events cut at each phase of pipeline and the final number of events that passed the pipeline with how many of them had mean FWHM less or more than $30$ days. ``Only Decline'' and ``Only Rise'' refer to objects that were observed only on the decline or the rise, and Cuts 1-4 to different phases of ClassPipe: 1. Linear fits, 2. Peak MJD of Gaussian fits, 3. Peak fluxes, 4. The average fluxes outside the event year. For more details see text.}
\label{tab:pipeline}
	\begin{tabular}{l | l l l l}
		\hline		
				&	DESY1 &	DESY2	&	DESY3	&	DESY4	\\
		\hline
		\textbf{Total}		&	4280	&	4161	&	3618	&	4021 \\
		\hline		
	 	Only Decline		&	302	&	235	&	242	&	252	\\
    	Only Rise		&	77	&	101	&	49	&	64	\\
        Failed Cut 1.			&	409	&	389	&	135	&	176 \\		
		Failed Cut 2.			&	774	&	808	&	862	&	1057	\\
		Failed Cut 3.			&	521	&	507	&	559	&	670	\\
		Failed Cut 4.			&	174	&	76	&	127	&	80	\\

		\hline
        \textbf{Passed ClassPipe}&	2023	&	2045	&	1644	&	1709	\\
		\hline
		FWHM$>30$ d &	1691	&	1783	&	1259	&	1359 \\	
        \hline	
        FWHM$<30$ d &	332	&	262	&	385	&	363 \\ 
        \hline
	\end{tabular}
\end{table}

\begin{table*}
\def\arraystretch{1.3}
\setlength\tabcolsep{4.5pt}
\centering
\fontsize{8}{10}\selectfont
\caption{Basic information of the host galaxies of the whole sample. The observing survey, the spectroscopic redshift of the host galaxy and the physical transient offset (in kpc)  has been given for the gold and silver sample transients, and the angular offset (in $\arcsec$), the Directional Light Radius \citep[DLR,][]{Sullivan2006} of the transient and the photometric redshift of the host galaxy \citep{Bonnett2016} for the whole sample. DLR is the ratio of the distance of the transient from the galaxy center and the half-light radius of that galaxy \citep[for more details see e.g.][]{Gupta2016}. Note that  $z_\mathrm{spec}$ for DES16E1bir was obtained from host galaxy spectral features present in the spectrum of the  transient (see Section \ref{sec:spectra}). The $^*$ symbols refers to three host galaxy candidates that are ``quasi-stellar": objects classified as stars based on their apparent shape and as such were not initially targeted for spectroscopic followup (but are now in the spectroscopy queue). Nine events in total appear to be hostless.}
\begin{tabular}{llccccc p{0.2em} llccccc}
\cline{1-7} \cline{9-15}
Name	&	Survey & Offset  & Offset & DLR & $z_\mathrm{spec}$ & $z_\mathrm{phot}$ &  & Name	&	Survey & Offset & Offset & DLR   & $z_\mathrm{spec}$ & $z_\mathrm{phot}$  \\
	&	 &  ($\arcsec$) & (kpc) &  &   & & & 	&	 &  ($\arcsec$) & (kpc) &   &   \\
\cline{1-7} \cline{9-15}
\multicolumn{15}{c}{\textbf{Gold Sample}} \\
DES13X1hav	& OzDES	& 0.381	& 2.51	& 0.90	& 0.58	& 0.64	&	& DES15S1fli	& OzDES	& 0.771	& 4.44	& 0.81	& 0.45	& 0.46	\\
DES13X3gms	& OzDES	& 0.879	& 6.09	& 1.13	& 0.65	& 0.64	&	& DES15S1fll	& OzDES	& 2.997	& 11.01	& 3.50	& 0.23	& 0.21	\\
DES14C3tvw	& ACES	& 3.703	& 26.46	& 3.71	& 0.70	& 0.70	&	& DES15X3mxf	& OzDES	& 1.726	& 9.81	& 3.14	& 0.44	& 0.43	\\
DES14S2anq	& SDSS	& 0.377	& 0.37	& 0.34	& 0.05	& 0.08	&	& DES16C1cbd	& OzDES	& 1.075	& 6.83	& 1.21	& 0.54	& 0.54	\\
DES14S2plb	& OzDES	& 1.506	& 3.26 	& 1.82	& 0.12	& 0.13	&	& DES16C2ggt	& PRIMUS& 1.462	& 6.66	& 1.98	& 0.31	& 0.35	\\
DES14S2pli	& OzDES	& 1.623	& 8.01	& 2.24	& 0.35	& 0.39	&	& DES16E1bir	& 	-	& 0.341	& 2.89	& 1.11	& 1.56	& 0.53	\\
DES14X3pkl	& OzDES	& 0.281	& 1.25	& 0.52	& 0.30	& 0.43	&	& DES16E2pv		& OzDES	& 1.081	& 7.85	& 1.84	& 0.73	& 0.82	\\
DES15C3lpq	& OzDES	& 0.323	& 2.18	& 0.44	& 0.61	& 0.62	& 	& DES16S1dxu	& OzDES	& 4.120	& 10.16	& 4.87	& 0.14	& 0.13	\\
DES15C3mgq	& OzDES	& 0.207	& 0.76	& 0.38	& 0.23	& 0.26	&	& DES16X1eho	& PRIMUS& 0.425	& 3.13	& 0.72	& 0.76	& 0.82	\\
DES15E2nqh	& OzDES	& 0.517	& 3.22	& 1.41	& 0.52	& 0.46	&	& DES16X3cxn	& OzDES	& 0.626	& 4.12	& 1.00	& 0.58	& 0.64	\\
\cline{1-7} \cline{9-15}
\multicolumn{15}{c}{\textbf{Silver Sample}} \\
DES13C1tgd	& OzDES	& 0.399	& 1.32	& 0.41	& 0.20	& 0.32	&	& DES15C3lzm	& ATLAS	& 0.511	& 2.43	& 0.47	& 0.33	& 0.30	\\
DES13C3bcok	& AAT 	& 0.791	& 3.91	& 0.39	& 0.35	& 0.30	&	& DES15C3nat	& OzDES	& 0.592	& 4.52	& 1.16	& 0.84	& 0.81	\\
DES13C3uig	& ACES	& 0.542	& 3.80	& 0.78	& 0.67	& 0.66	&	& DES15C3opk	& OzDES	& 0.523	& 3.41	& 0.76	& 0.57	& 0.59	\\
DES13E2lpk	& OzDES	& 0.799	& 4.77	& 0.79	& 0.48	& 0.38	&	& DES15C3opp	& OzDES	& 0.429	& 2.44	& 0.83	& 0.44	& 0.37	\\
DES13X3npb	& OzDES	& 0.173	& 1.06	& 0.16	& 0.50	& 0.37	&	& DES15X2ead	& OzDES	& 0.545	& 2.00	& 0.65	& 0.23	& 0.25	\\
DES13X3nyg	& OzDES	& 0.410	& 2.95	& 0.75	& 0.71	& 0.68	&	& DES16C3axz	& AAT 	& 0.336	& 1.23	& 0.44	& 0.23	& 0.21	\\
DES13X3pby	& OzDES	& 0.337	& 2.67	& 0.75	& 0.96	& 0.41	&	& DES16C3gin	& OzDES	& 1.606	& 7.93	& 1.87	& 0.35	& 0.41	\\
DES14X1bnh	& OzDES	& 0.452	& 3.44	& 0.68	& 0.83	& 0.80	&	& DES16X3ega	& GAMA	& 2.219	& 8.93	& 2.17	& 0.26	& 0.19	\\
DES15C2eal	& OzDES	& 2.443	& 8.68	& 5.50	& 0.22	& 0.26	\\
\cline{1-7} \cline{9-15}
\multicolumn{15}{c}
{\textbf{Bronze Sample}} \\
DES13C1acmt		& -	& 0.571	& -	& 0.88	& -	& 0.61	& 	& DES15C3mfu	& -	& -		& -	& -		& -	& -	 \\
DES13C3abtt$^*$	& -	& 2.163	& -	& 3.58	& -	& 0.58	& 	& DES15C3pbi$^*$&-	& 0.058 & -	& 0.16	& -	& 1.06	\\
DES13C3asvu$^*$	& -	& 0.331	& -	& 0.55	& -	& 0.83	& 	& DES15E2lmq	& -	& 0.055	& -	& 0.19	& -	& 1.14	\\
DES13C3avkj		& -	& 0.185	& -	& 0.39	& -	& 0.96	& 	& DES15X3atd	& -	& 0.503 & -	& 0.92	& -	& 1.19	\\
DES13C3nxi		& -	& 0.551	& -	& 1.33	& -	& 0.60	& 	& DES15X3kyt	& -	& 0.313	& -	& 0.81	& -	& 1.50	\\
DES13C3smn		& -	& 0.132	& -	& 0.41	& -	& 1.50	& 	& DES15X3nlv	& -	& 0.197	& -	& 0.70	& -	& 1.07	\\
DES13X2oyb		& -	& 0.411	& -	& 0.78	& -	& 1.27	& 	& DES15X3oma	& -	& 0.341	& -	& 1.11	& -	& 0.67	\\
DES13X3aakf		& -	& 0.734	& -	& 1.85	& -	& 1.18	& 	& DES16C1bde	& -	& 0.655	& -	& 0.973	& -	& 0.97	\\
DES13X3afjd		& -	& -		& -	& -		& -	& -		& 	& DES16C2grk	& -	& 2.730	& -	& 4.34	& -	& 0.36	\\
DES13X3alnb		& -	& 0.147	& -	& 0.44	& -	& 0.68	& 	& DES16C3auv	& -	& 0.057	& -	& 0.13	& -	& 0.98	\\
DES13X3kgm		& -	& -		& -	& -		& -	& -		& 	& DES16C3cdd	& -	& 0.213	& -	& 0.45	& -	& 1.40	\\
DES14C1jnd		& -	& 0.186	& -	& 0.55	& -	& 0.43	&  	& DES16S2fqu	& -	& -		& -	& -		& -	& -	 	\\
DES14C3htq		& -	& -		& -	& -		& -	& -		& 	& DES16X1ddm	& -	& 1.921	& -	& 3.15	& -	& 0.38	\\
DES14E2bfx		& -	& -		& -	& -		& -	& -		&	& DES16X2bke	& -	& 1.615	& -	& 4.21	& -	& 0.90	\\
DES14E2xsm		& -	& 0.507	& -	& 1.16	& -	& 0.66	& 	& DES16X3ddi	& -	& 1.075	& -	& 2.30	& -	& 0.91	\\
DES14X1qzg		& -	& -		& -	& -		& -	& -		& 	& DES16X3erw	& -	& 0.455	& -	& 0.76	& -	& 0.40	\\
DES14X3pko		& -	& -		& -	& -		& -	& -		& 	& DES16X3wt		& -	& - 	& -	& -		& -	& -	 \\
DES15C3edw		& -	& 1.941	& -	& 2.63	& -	& 0.33	& 	& \\

\cline{1-7} \cline{9-15}
\end{tabular}
\label{tab:z_sources}
\end{table*}

\subsection{Selection Criteria for Rapidly Evolving Transients}
\begin{figure*}
\includegraphics[width=\textwidth]{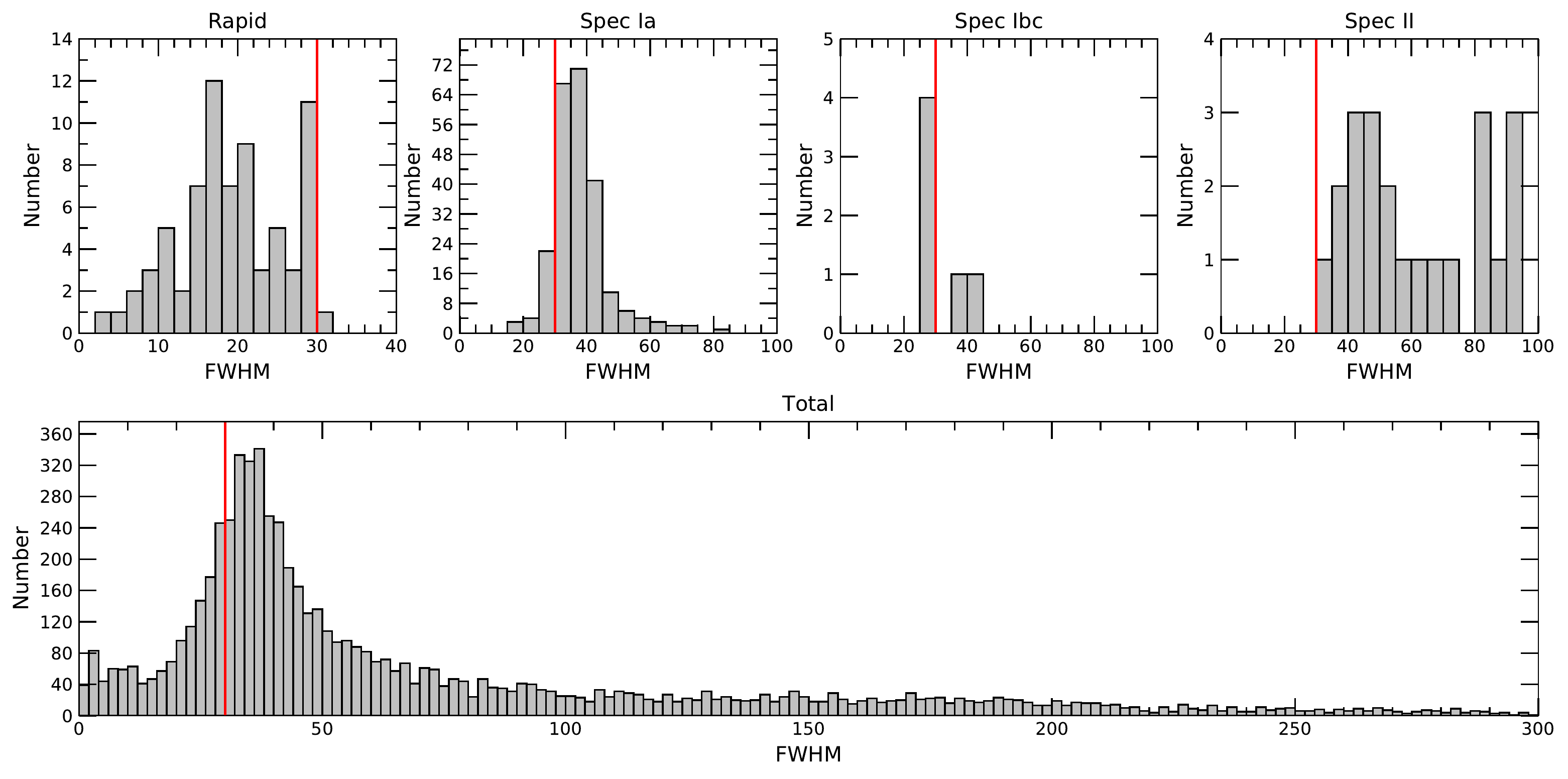}

\caption{Mean FWHM distribution of the rapidly evolving transients in comparison with distributions of spectroscopically confirmed samples of Type Ia, Ibc and II SNe (top) and FWHM distribution of all events that passed ClassPipe (see Table \ref{tab:pipeline}). Red line refers to the cutoff limit of FWHM $=30$ d for the rapidly evolving transients. Note that one rapid event, DES16C1cbd, has FWHM $>30$ d. This event was found as part of different analysis and was included in our sample due to apparent similarity of the light curve evolution with the rest of our sample. All events found above FWHM $=300$ d are ignored in the bottom panel. }
\label{fig:fwhm_hist}
\end{figure*}
Rapidly evolving transients previously presented in the literature evolve on timescales that are considerably faster than any other type of SNe.  Thus to search for these objects in our sample of DES-SN transients, we expect these events should have a small FWHM. The cutoff limit was then chosen based on the FWHM distributions of spectroscopically confirmed Type Ia, Ibc and II SNe from first three years of DES-SN shown in the top panel of Figure \ref{fig:fwhm_hist}. As FWHM is only a rough measurement of duration we wanted to look at intervals of FWHM that partially overlap with distributions of some traditional SNe, in order not to lose interesting events. Thus, we chose FWHM below 30 days which overlaps only with a small number of Type Ia and Ibc.  The selection also focuses on the faster end of all events that passed ClassPipe as seen in the FWHM distribution plotted in the bottom of Figure \ref{fig:fwhm_hist}.

Our sample of rapidly evolving transients was then found by visually inspecting the light curves of all $\approx1350$ events that had mean FWHM less than 30 days as shown in Table \ref{tab:pipeline}. Most of these candidates were ultimately rejected as the small value of FWHM was driven by a single, possibly low signal-to-noise ratio, data point above otherwise constant background level. However, close to FWHM $=30$ d we also excluded traditional types of SNe based on the light curve evolution.

As a result of this stage of visual inspection, we confirmed a sample of 72 transients with robust detections satisfying all the above selection criteria. These transients are characterized by a rapid rise to peak brightness in $t_\mathrm{rise}\lesssim10$ d and a subsequent fading occurring over the following $t_\mathrm{decl}\lesssim30$ d in the observer frame. However, as we were solely looking for rapidly evolving events, we might have missed events that have similar light curve evolution but last for a slightly longer period of time in observer frame and hence have larger FWHM. The longer duration can either be caused by time dilation of bright, high redshift events or intrinsic dispersion caused by the underlying physical mechanism. One such event is DES16C1cbd, with FWHM $=30.9$ d, which was discovered during visual inspection done as a part of different analysis. The transient was included in our sample due to apparent similarity of the light curve evolution with the rest of the sample, and hence is the only transient in our sample with FWHM $>30$.  We present the detailed analysis of the found 72 objects in the Sections that follow.

\subsection{Spectroscopic Observations}
Many objects in our sample had spectroscopic observations of the host galaxy of the transient object, and in a few cases a spectrum was taken of the transient event itself. In Table \ref{tab:z_sources} we present the host galaxy information including the spectroscopic instruments that were used to observe the host galaxy and host galaxy redshifts. These redshifts are collectively composed from the ``Global Redshift Catalog" compiled by OzDES \citep{yuan15}, which contains redshifts obtained by OzDES itself as well as other literature spectroscopy campaigns conducted in the DES-SN fields.  The vast majority of redshifts come from OzDES observations on the 4m AAT telescope in Australia, with some additions from other AAT programs (e.g., GAMA), as well as some redshifts from SDSS \citep{york00}, PRIMUS \citep{coil11}, ACES \citep{cooper12}, and ATLAS \citep{norris06}.  The full details of the observational operations of OzDES are described in \citet{yuan15} and updated in \citet{Childress2017}.

Spectra of three rapid transients were obtained as part of the multi-facility spectroscopic followup program for DES-SN.  The full description of this program will be described in D'Andrea et al. (2018, in prep.). A spectrum of DES14S2anq was obtained with the 4m Anglo-Australian Telescope in Australia as part of the OzDES programme.  DES15C3opk was spectroscopically observed with the Low Dispersion Survey Spectrograph (LDSS3) on the Magellan Clay Telescope at Las Campanas in Chile under RAISINS2 program.  Finally, we obtained a spectrum of DES16E1bir with X-SHOOTER \citep{Vernet2011} on the 8m Very Large Telescope (VLT) at Paranal observatory in Chile. Spectroscopic redshifts for DES15C3opk and DES16E1bir were obtained from the corresponding spectra (see Section \ref{sec:hostgal}).

\section{Sample Overview}
\label{sec:sample}
\begin{table*}
\def\arraystretch{1.1}%
\setlength\tabcolsep{6pt}
\centering
\fontsize{8}{10}\selectfont
\caption{The sample of 72 rapidly evolving transients including 37 with spectroscopic redshift. Given redshifts are from the spectra of the host galaxies except for DES15C3opk and DES16E1bir for which they were obtained from host galaxy spectral features present in spectra of the events (see section \ref{sec:spectra}). The Milky way color excesses are taken from \citet{Schlafly2011}.}
\begin{tabular}{l c c c c p{0.5em} l c c c c}
\cline{1-5} \cline{7-11}
		Name	&	R.A.	&	Decl.	&	$z_\mathrm{spec}$ & $E_\mathrm{B-V}$	& & Name	&	R.A.	&	Decl.	&	$z_\mathrm{spec}$ & $E_\mathrm{B-V}$	\\
                  &	(J2000)	&	(J2000)	&		 & 					& & 		&	(J2000)	&	(J2000)	&	 & 	\\
\cline{1-5} \cline{7-11}
\multicolumn{11}{c}{\textbf{Gold Sample}} \\
DES13X1hav	& 02:20:07.80	& -05:06:36.53	& 0.58	& 0.0185 & & DES15S1fli	& 02:52:45.15	& -00:53:10.21	& 0.45	& 0.0675\\
DES13X3gms	& 02:23:12.27	& -04:29:38.35	& 0.65	& 0.0246 & & DES15S1fll	& 02:51:09.36	& -00:11:48.71	& 0.23	& 0.0611\\
DES14C3tvw	& 03:33:17.61	& -27:54:23.92	& 0.70	& 0.0058 & & DES15X3mxf	& 02:26:57.72	& -05:14:22.81	& 0.44	& 0.0237\\
DES14S2anq	& 02:45:06.67	& -00:44:42.77	& 0.05	& 0.0294 & & DES16C1cbd	& 03:39:25.97	& -27:40:20.37	& 0.54	& 0.0100\\
DES14S2plb	& 02:47:25.62	& -01:37:06.91	& 0.12	& 0.0369 & & DES16C2ggt	& 03:35:33.88	& -29:13:29.33	& 0.31	& 0.0089\\
DES14S2pli	& 02:44:54.76	& -01:05:52.74	& 0.35	& 0.0256 & & DES16E1bir	& 00:30:58.64	& -42:58:37.18	& 1.56 	& 0.0064\\
DES14X3pkl	& 02:28:50.64	& -04:48:26.44	& 0.30	& 0.0332 & & DES16E2pv	& 00:36:50.19	& -43:31:40.16	& 0.73	& 0.0059\\
DES15C3lpq	& 03:30:50.89	& -28:36:47.08	& 0.61	& 0.0078 & & DES16S1dxu	& 02:50:43.53	& -00:42:33.29	& 0.14	& 0.0486\\
DES15C3mgq	& 03:31:04.56	& -28:12:31.74	& 0.23	& 0.0080 & & DES16X1eho	& 02:21:22.87	& -04:31:32.64	& 0.76	& 0.0229\\
DES15E2nqh	& 00:38:55.59	& -43:05:13.14	& 0.52	& 0.0076 & & DES16X3cxn	& 02:27:19.32	& -04:57:04.27	& 0.58	& 0.0234\\
		\cline{1-5} \cline{7-11}
      \multicolumn{11}{c}{\textbf{Silver Sample}} \\
DES13C1tgd	& 03:36:15.42	& -27:38:19.07	& 0.20	& 0.0105 & & DES15C3lzm	& 03:28:41.86	& -28:13:54.96	& 0.33	& 0.0063\\
DES13C3bcok	& 03:32:06.47	& -28:37:29.70	& 0.35	& 0.0081 & & DES15C3nat	& 03:31:32.44	& -28:43:25.06	& 0.84	& 0.0086\\
DES13C3uig	& 03:31:46.55	& -27:35:07.96	& 0.67	& 0.0073 & & DES15C3opk	& 03:26:38.76	& -28:20:50.12	& 0.57	& 0.0119\\
DES13E2lpk	& 00:40:23.80	& -43:32:19.74	& 0.48	& 0.0058 & & DES15C3opp	& 03:26:57.53	& -28:06:53.61	& 0.44	& 0.0084\\
DES13X3npb	& 02:26:34.11	& -04:08:01.96	& 0.50	& 0.0242 & & DES15X2ead	& 02:25:57.38	& -06:27:04.78	& 0.23	& 0.0289\\
DES13X3nyg	& 02:27:58.17	& -03:54:48.05	& 0.71	& 0.0233 & & DES16C3axz	& 03:31:14.15	& -28:40:00.25	& 0.23	& 0.0082\\
DES13X3pby	& 02:25:19.98	& -05:18:50.58	& 0.96	& 0.0223 & & DES16C3gin	& 03:31:03.06	& -28:17:30.98	& 0.35 	& 0.0082\\
DES14X1bnh	& 02:14:59.79	& -04:47:33.32	& 0.83	& 0.0173 & & DES16X3ega	& 02:28:23.71	& -04:46:36.18	& 0.26	& 0.0308\\
DES15C2eal	& 03:36:14.68	& -29:13:49.32	&  0.22	& 0.0086\\
		\cline{1-5} \cline{7-11}
       \multicolumn{11}{c}{\textbf{Bronze Sample}} \\
DES13C1acmt	& 03:37:18.99	& -26:50:00.99	&  - 	& 0.0101 & & DES15C3mfu	& 03:28:36.08	& -28:44:20.00	&  - 	& 0.0085\\
DES13C3abtt	& 03:30:28.91	& -28:09:42.12	&  - 	& 0.0065 & & DES15C3pbi	& 03:28:56.68	& -28:00:07.98	&  - 	& 0.0073\\
DES13C3asvu	& 03:31:20.82	& -27:21:38.89	&  - 	& 0.0086 & & DES15E2lmq	& 00:38:28.82	& -43:59:14.04	&  - 	& 0.0055\\
DES13C3avkj	& 03:27:52.97	& -27:31:40.86	&  - 	& 0.0078 & & DES15X3atd	& 02:23:21.64	& -04:17:28.95	&  - 	& 0.0230\\
DES13C3nxi	& 03:27:51.22	& -28:21:26.21	&  - 	& 0.0071 & & DES15X3kyt	& 02:25:05.98	& -05:24:39.65	&  - 	& 0.0229\\
DES13C3smn	& 03:27:53.08	& -28:05:00.93	&  - 	& 0.0080 & & DES15X3nlv	& 02:24:10.70	& -05:01:38.47	&  - 	& 0.0237\\
DES13X2oyb	& 02:21:14.55	& -05:40:44.03	&  - 	& 0.0212 & & DES15X3oma	& 02:26:59.07	& -05:06:37.70	&  - 	& 0.0239\\
DES13X3aakf	& 02:22:50.84	& -04:41:57.01	&  - 	& 0.0217 & & DES16C1bde	& 03:37:12.34	& -26:45:29.93	&  - 	& 0.0105\\
DES13X3afjd	& 02:28:00.31	& -04:34:59.39	&  - 	& 0.0248 & & DES16C2grk	& 03:40:16.35	& -29:17:20.18	&  - 	& 0.0112\\
DES13X3alnb	& 02:28:44.39	& -05:08:32.47	&  - 	& 0.0249 & & DES16C3auv	& 03:27:36.51	& -27:35:27.33	&  - 	& 0.0077\\
DES13X3kgm	& 02:26:00.92	& -04:51:59.29	&  - 	& 0.0231 & & DES16C3cdd	& 03:29:42.53	& -27:08:35.49	&  - 	& 0.0081\\
DES14C1jnd	& 03:37:24.36	& -27:29:35.03	&  - 	& 0.0095 & & DES16S2fqu	& 02:47:05.94	& -00:20:50.40	&  - 	& 0.0313\\
DES14C3htq	& 03:30:17.34	& -27:41:46.61	&  - 	& 0.0071 & & DES16X1ddm	& 02:15:18.88	& -04:21:52.07	&  - 	& 0.0183\\
DES14E2bfx	& 00:42:40.58	& -44:25:05.62	&  - 	& 0.0057 & & DES16X2bke	& 02:24:02.30	& -07:16:17.87	&  - 	& 0.0256\\
DES14E2xsm	& 00:38:42.27	& -43:35:14.11	&  - 	& 0.0067 & & DES16X3ddi	& 02:21:45.39	& -04:41:08.95	&  - 	& 0.0215\\
DES14X1qzg	& 02:19:43.65	& -05:26:30.87	&  - 	& 0.0197 & & DES16X3erw	& 02:24:49.31	& -04:30:51.45	&  - 	& 0.0205\\
DES14X3pko	& 02:27:37.95	& -03:41:44.27	&  - 	& 0.0268 & & DES16X3wt	& 02:24:56.98	& -05:32:42.19	&  - 	& 0.0224\\
DES15C3edw	& 03:30:12.59	& -27:42:34.93	&  - 	& 0.0071 & & \\
\cline{1-5} \cline{7-11}
\end{tabular}
\label{tab:sample}
\end{table*}

\begin{figure*}
\includegraphics[width=\textwidth]{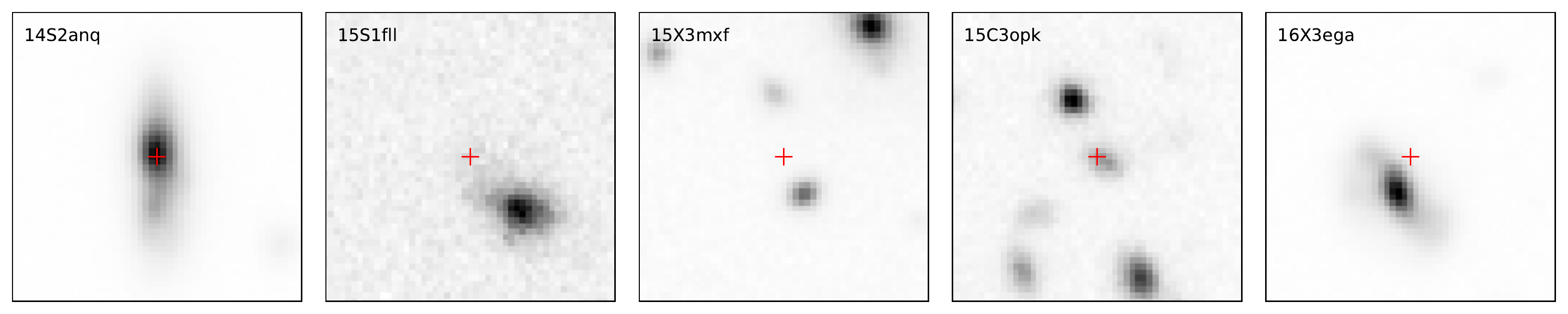}
\caption{Explosion environments for five gold and silver sample transients. Transient explosion sites are marked with red crosses. Each square box is roughly $13.5\arcsec$ on a side.}
\label{fig:stamps}
\end{figure*}

Basic information about the 72 rapid transients and their host galaxies is presented in Table \ref{tab:sample}, and we show example environments for five transients from our sample Figure~\ref{fig:stamps}. In this sample there are no events with multiple host candidates of similar host-transient separation that could lead to an ambiguous host association. For the remainder of this paper, we split our sample into three groups based on the quality of the light curves (specifically, the number of bands and epochs in which the transient is detected at $3\sigma$ in image subtractions) and availability of a redshift.  These groups will be referred to as ``gold", ``silver" and ``bronze", and their selection is described as follows.

\subsection{Gold Sample}

\begin{figure*}
\includegraphics[width=\textwidth]{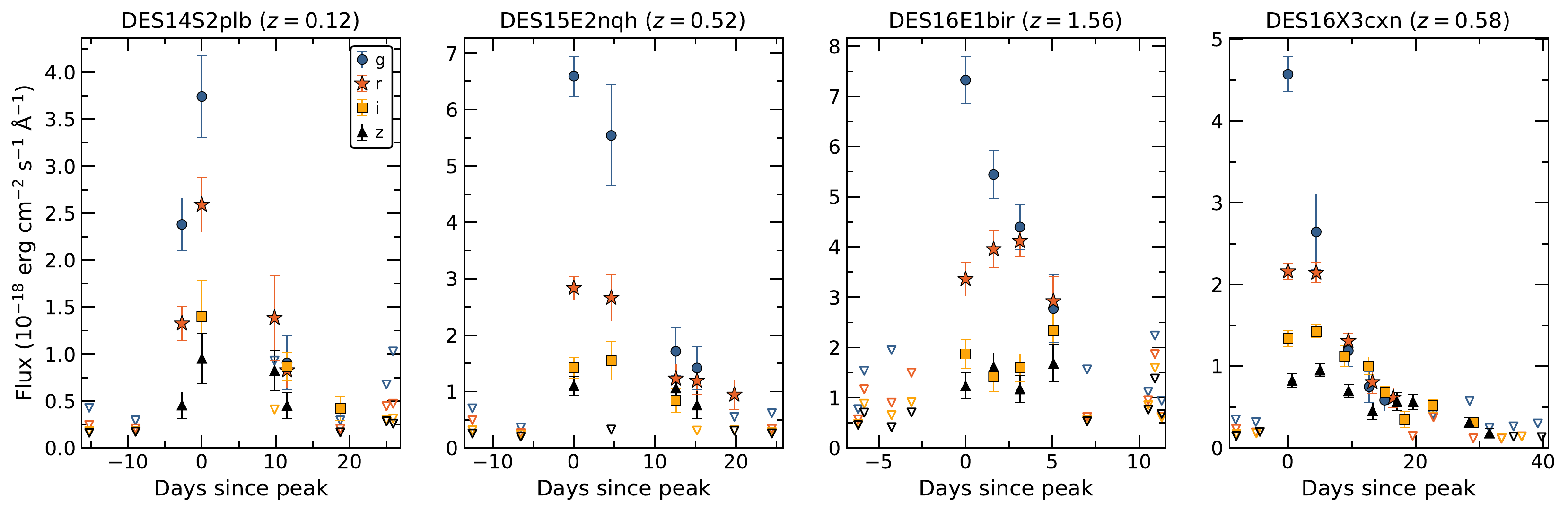}
\caption{Rest frame light curves since peak in $g$ band of four gold sample transients. Open triangles represent 1$\sigma$ error of data points below 3$\sigma$ detection. }
\label{fig:gold_lcs}
\end{figure*}

\begin{figure*}
\includegraphics[width=\textwidth]{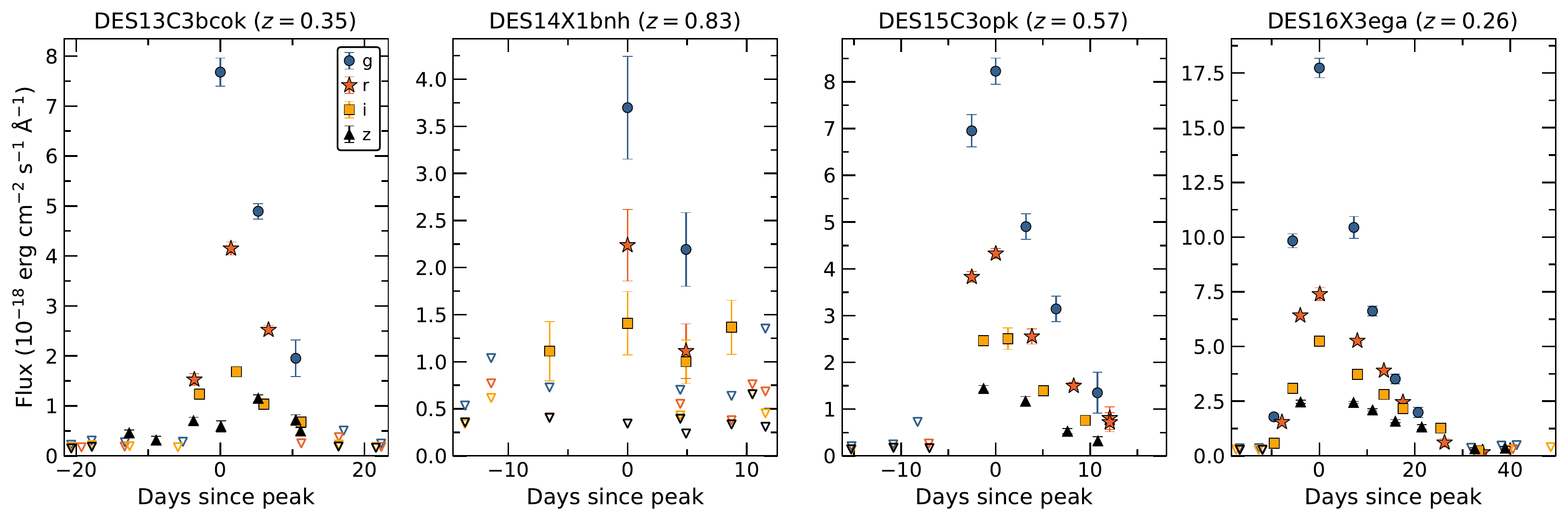}
\caption{Same as Figure \ref{fig:gold_lcs} for four silver sample transients.}
\label{fig:silver_lcs}
\end{figure*}

\begin{figure*}
\includegraphics[width=\textwidth]{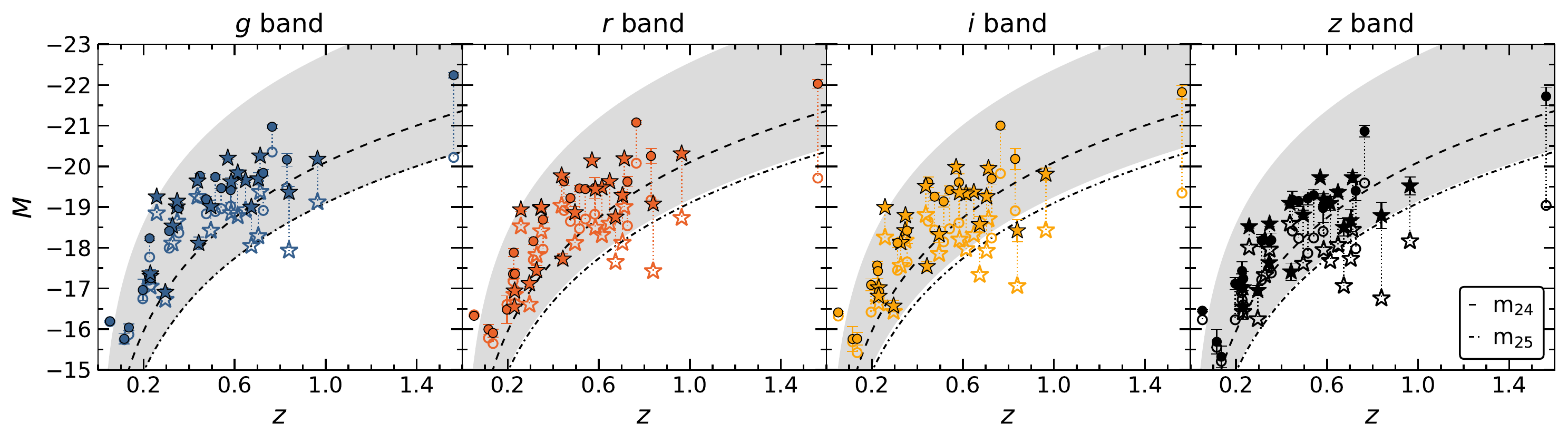}
\caption{Absolute magnitude at peak plotted against redshift, derived in each band using the epoch closest to the observed $g$ band peak. Objects found in DES-SN deep and shallow fields are marked with stars and circles, respectively. The open markers represent corresponding values that have been $k$-corrected based on the best blackbody fits at peak (see Section \ref{subsec:bb}). Dashed and dot-dashed lines correspond to apparent magnitudes of 24 and 25, respectively. $k$-corrected values have not been plotted for DES13C3bcok, DES15C3opk, DES15C3opp and DES16C3axz for which blackbody fits at peak were based on two bands and are therefore dubious (see Table \ref{tab:bb_peak}). Grey band underneath represents the region where bronze sample transients would be based on their apparent peak magnitudes. Only event at $z>1$ is DES16E1bir at $z=1.56$.}
\label{fig:M_z}
\end{figure*}

\begin{figure*}
\includegraphics[width=\textwidth]{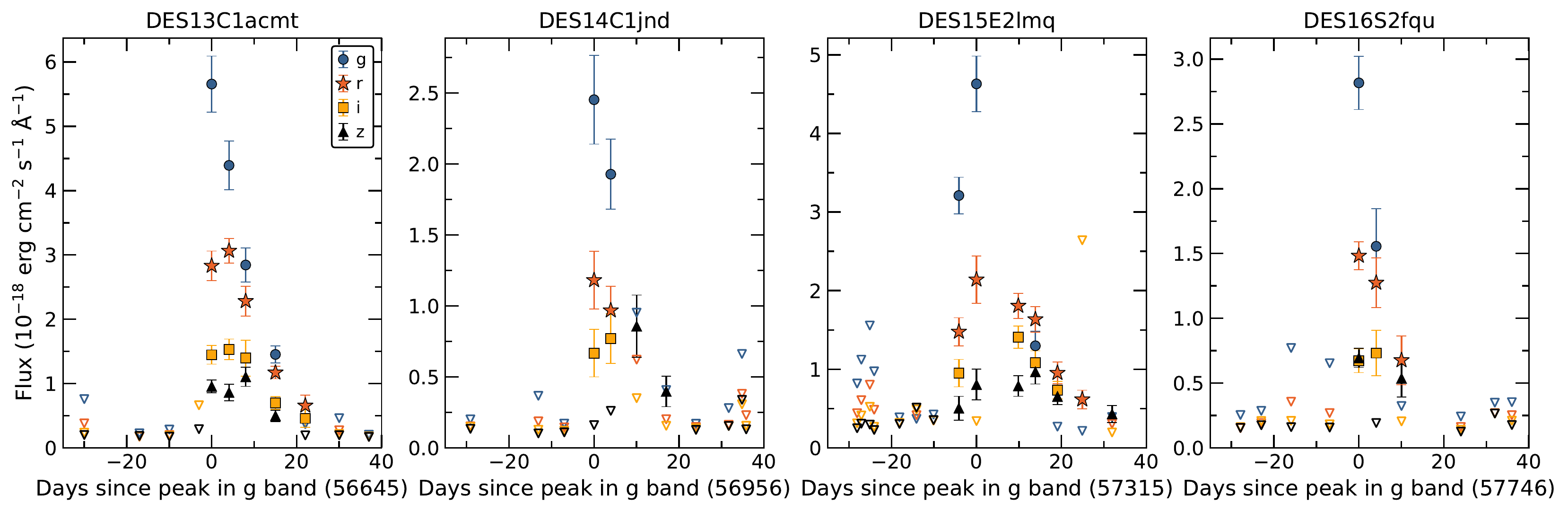}
\caption{Observer frame light curves of four bronze sample transients. Open triangles represent 1$\sigma$ error of data points below 3$\sigma$ detection.}
\label{fig:bronze_lcs}
\end{figure*}

Our gold sample consists of 20 transients.  These objects have known host galaxy redshifts, allowing us to constrain their true luminosities, but they also had to pass cuts based on the quality of the photometric data.  Later in this paper we perform blackbody fits on the photometric data of our transients on every 1.5 rest frame day ``epochs" with observations in more than one band (achieved by checking 1.5 d window surrounding each data point to find the best combinations). The value was chosen so blackbody fits were also possible for events in DES-SN deep fields, where bands are not always observed during the same night. Using the 1.5 d epochs, we classified transient with a redshift as gold if it had at least three epochs with data in three or four bands (which makes the blackbody fits more reliable). If this was not the case, transient was classified as silver.  For these events we analyze their temperature and radius evolution. Example light curves of four gold sample objects are presented in Figure~\ref{fig:gold_lcs}.

\subsection{Silver Sample}

We define the silver sample as those objects with redshifts but which do not fulfill the photometric requirements for the gold sample -- this silver sample consists of a total of 17 objects. They have either less than three epochs altogether or have several epochs but less than three with data in more than two bands. Events in the silver sample are mostly in the DES-SN deep fields, where different bands are often observed on different days. This allows accurate measurement of the photometric evolution of the transient (e.g., rise and decline timescales), but precludes measurement of the evolution of physical properties like temperature and radius. Example light curves of four silver sample transients are presented in Figure \ref{fig:silver_lcs}.

Based on the 37 objects in our gold and silver samples, the rapidly evolving transients span a wide range of redshifts (0.05 to 1.5) and absolute magnitudes (-15.75 to -22.25). This can be clearly seen in Figure~\ref{fig:M_z} where we have plotted the absolute magnitude at the observed epoch closest to peak as measured in $g$ band in all four bands against the redshift. The majority of the observed transients are found between redshifts $0.2$ and $0.8$ with only two above and three below. Above redshift $z=0.8$, key emission features such as [OII] $\lambda$3727 Å are redshifted into wavelength ranges heavily contaminated by night sky lines, making redshift success particularly difficult for OzDES \citep{Childress2017}. In this plot we have also given the roughly $k$-corrected values based on the best blackbody fits at peak (see Section \ref{subsec:bb}). Light curves of all 37 gold and silver sample transients are presented in Appendix \ref{app:gold_lc}.

\subsection{Bronze Sample}

\begin{figure}
\includegraphics[width=0.485\textwidth]{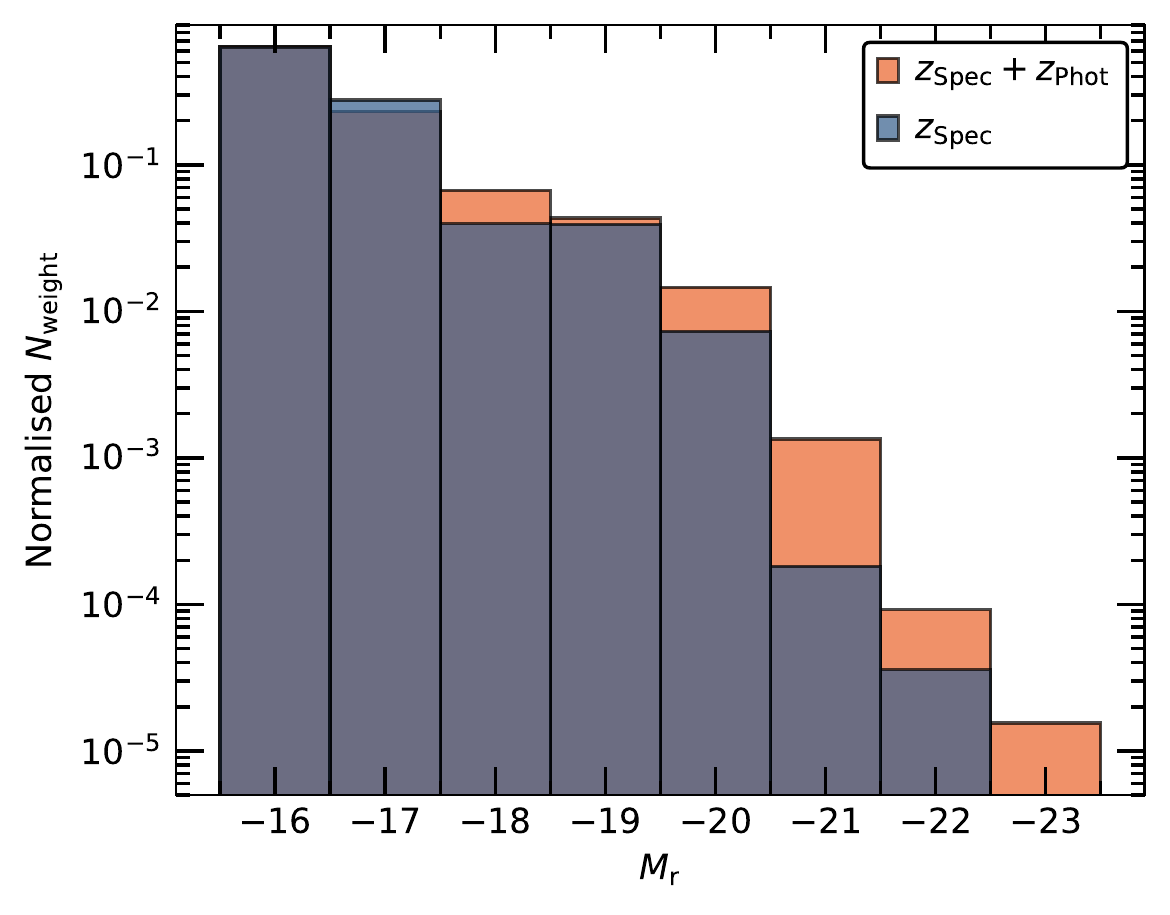}
\caption{Normalized 1/V$_\mathrm{max}$ luminosity distribution for 37 events with spectroscopic redshifts (Blue) and for all 63 events associated with a host galaxy including the bronze sample transients (Orange). Distributions have been normalized so that area under the curve equals unity.}
\label{fig:N_Vmax}
\end{figure}

Our bronze sample consists of the 35 transients currently without redshifts.  Example observer frame light curves for four bronze sample events are plotted in Figure~\ref{fig:bronze_lcs} and for the whole sample in Appendix \ref{app:bronze_lc}. These events have light curves similar in shape to the ones in the gold and silver samples, but lack a spectroscopic redshift allowing a precise determination of the absolute brightness. Many of the objects in our bronze sample have reliable host galaxy identification where the host has a photometric redshift estimate, which we report in Table \ref{tab:z_sources} for all available hosts (including those with a spectroscopic redshift). These $z_\mathrm{phot}$ values allow us to make crude estimates of distances to the bronze sample events, but due to the uncertainty on these distances the majority of this manuscript will focus on the analysis of the gold and silver sample transients. 

\begin{figure*}
\includegraphics[width=0.98\textwidth]{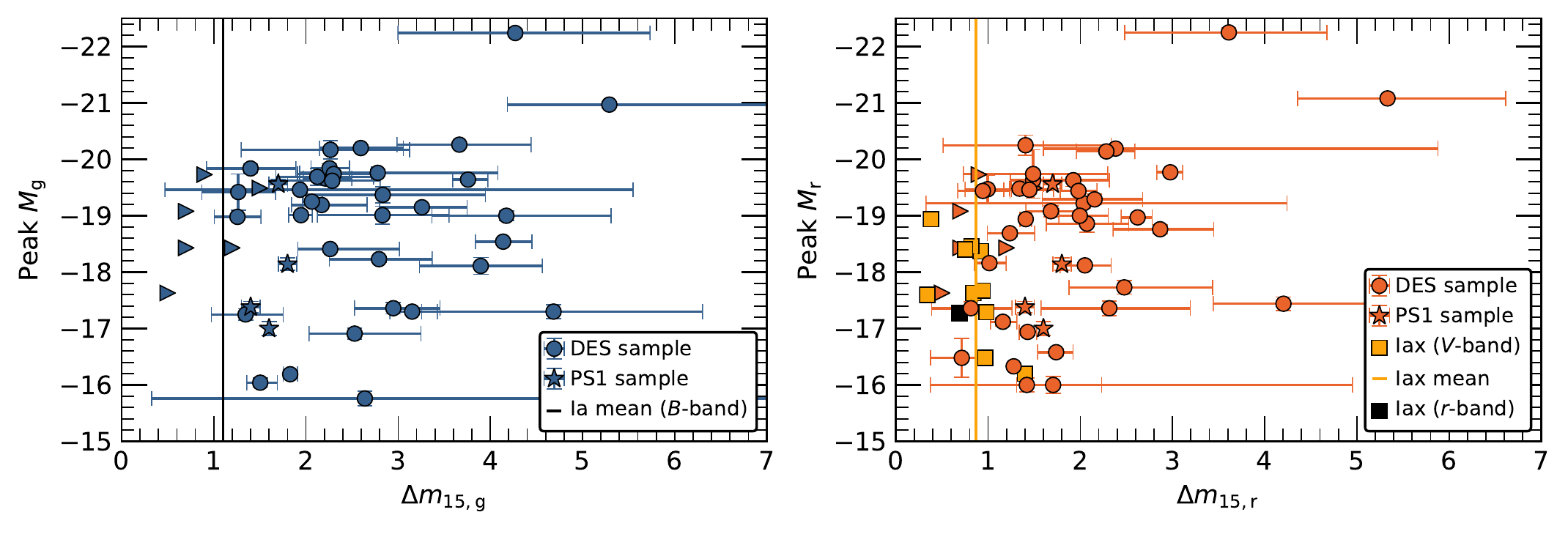}
\caption{$\Delta m_\mathrm{15}$ of gold and silver sample transients in $g$ (left) and $r$ (right) bands in comparison with other types of SNe. The $\Delta m_\mathrm{15}$ for the prototypical normal Type Ia SN 2011fe \citep{Pereira2013} in $B$ band is considerably smaller than $\Delta m_\mathrm{15,g}$ for all of our events.  $\Delta m_\mathrm{15}$ for Iax SNe in $V$ band \citep{Foley2013} partly overlap with the slower end of our events in $r$ band. However, as $V$ band is bluer than $r$ one would expect it to have a higher value in comparison. Additionally, $\Delta m_\mathrm{15}$ for Iax SN 2015H in $r$ band \citep{Magee2016} is smaller than most measured in $V$ band. The values of $\Delta m_{15}$ coincide well with the PAN-STARRS1 (PS1) events from \citet{Drout2014}. Right pointing triangles refer to PS1 events that have only lower limit on $\Delta m_\mathrm{15}$. $\Delta m_\mathrm{15}$ for our events has been estimated based on exponential fit on the decline described in Section \ref{subsec:t_rise_dec}.}
\label{fig:dm15_comp}
\end{figure*}

\begin{figure}
\includegraphics[width=0.48\textwidth]{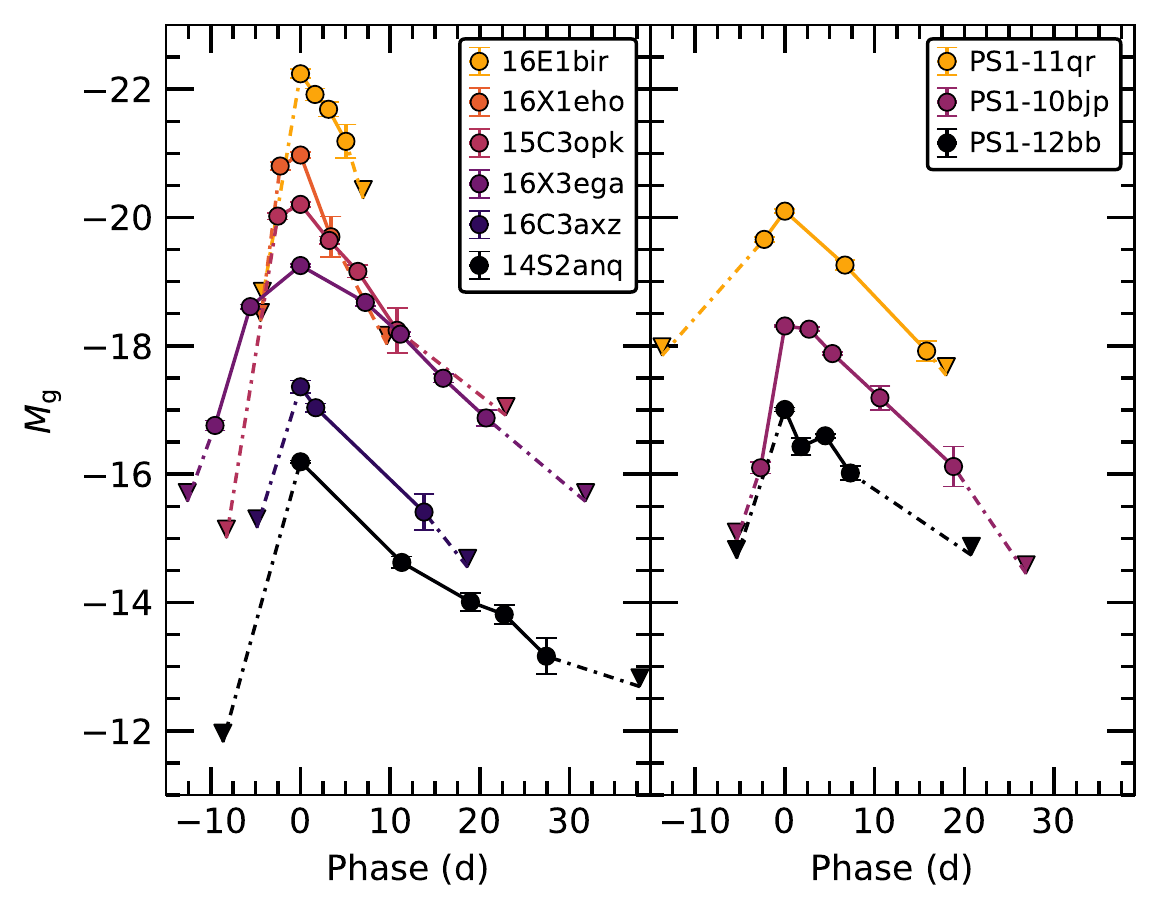}
\caption{Example $g$-band light curves for six gold and silver sample transients, in comparison with three events from \citet{Drout2014} (PS1-11qr at $z=0.32$, PS1-10bjp at $z=0.11$, PS1-12bb at $z=0.10$). We note that the transients are found at wide range of redshifts and thus sample a range of rest-frame wavelengths.}
\label{fig:drout_comp}
\end{figure}

In Figure~\ref{fig:N_Vmax} we present a $1/V_\mathrm{max}$ luminosity distribution (i.e. distribution weighted by absolute volume individual transient can be detected in.) for two subsets of transients: the 37 events with spectroscopic redshifts (shown in blue), and all 63 transients with redshift estimates (shown in orange). The luminosity function peaks at low luminosities, but these objects are only accessible to DES-SN in the low redshift Universe. However, the high luminosity tail is visible out to $z\gtrsim1$. The distribution for all events with associated host galaxy shows a slight shift towards higher luminosities in comparison with the spectroscopic sample.  

Using a similar technique to $1/V_\mathrm{max}$ luminosity distribution above, we can estimate a rough rate based on the sample of 37 events with spectroscopic redshifts. We found a rate of $\gtrsim10^{-6}$ events Mpc$^{-3}$ yr$^{-1}$, which is $\approx$1.5\% of the volumetric CCSN rate \citep{Li2011a}. We note this is a lower limit of the rate, as it was estimated with the assumption of complete sample (down to the limiting magnitudes of the DES-SN search fields). Hence, the rate is smaller than, but consistent with, the rate of $4.8-8\cdot10^{-6}$ events Mpc$^{-3}$ yr$^{-1}$ calculated by \citet{Drout2014} for the Pan-STARRS1 sample of rapidly evolving transients. The full rate analysis for our DES-SN sample of rapidly evolving transients will be featured in a forthcoming paper.

\section{Photometric Properties}
\label{sec:photo}

The majority of the gold and silver sample transients have rapid photometric evolution ($t_\mathrm{1/2}\lesssim 12$ d in rest-frame), which increases the number of optical transients with such short timescales by more than factor of two. Our sample also consists of transients in a very wide range of $-15.75>M_\mathrm{g}>-22.25$. In this paper we refer to observed brightest data point in g band as the peak of the event, while rise and decline times for a given band are calculated with respect to the date of observed brightest data point in that band.

In Figure \ref{fig:dm15_comp} we show that $\Delta m_{15}$ \citep[number of magnitudes events decline in first 15 days after the peak,][]{Phillips1993} of our transients are similar with $\Delta m_{15}$ of the PAN-STARRS sample \citep{Drout2014}, but significantly larger when compared with Ia and Iax SNe. The light curves evolution is also similar to PAN-STARRS events, as can be seen in Figure \ref{fig:drout_comp} where we have plotted example $g$ band light curve in comparison with three events found in \citet{Drout2014}. However, we want to emphasize that plotted events are found at wide range of redshifts and thus will sample a range of rest-frame wavelengths. In this section we will discuss photometric properties of the gold and silver sample transients.

\subsection{Rise and Decline Timescales}
\label{subsec:t_rise_dec}
\begin{table}
\def\arraystretch{1.2}%
\setlength\tabcolsep{4pt}
\fontsize{8}{10}\selectfont
\caption{Light curve parameters of gold and silver sample transients in $g$ band. $t_\mathrm{peak}$ is defined as the date of the maximum observed $g$ band flux, while rise time $t_\mathrm{rise}$, decline time $t_\mathrm{decline}$ and time above half maximum $t_\mathrm{1/2}$ are given in rest frame. $t_\mathrm{rise}$ is defined to be the time between last non-detection and observed peak and $t_\mathrm{decline}$ the time exponential fit takes to decline to one tenth of the observed peak flux. Errors for $t_\mathrm{decline}$ are given in 1$\sigma$ confidence, but no errors are given for $t_\mathrm{rise}$ (and $t_\mathrm{1/2}$) as the values are effectively upper limits. $t_\mathrm{decline}$ and $t_\mathrm{1/2}$ has not been given for DES13C1tgd and DES13X3pby as they have only one detection in $g$ band.}
\begin{adjustbox}{max width=0.49\textwidth, center=0.45\textwidth}
\centering
\begin{tabular}{l c c c c c}
\hline
Name	&	$t_\mathrm{peak}$	&	$M_\textrm{peak}$	&	$t_\mathrm{rise}$	& $t_\mathrm{decline}$& $t_\mathrm{1/2}$ \\
		&	(MJD)				&	(days)				&	(days)				& (days)		& (days) \\
\hline     
\multicolumn{6}{c}{\textbf{Gold Sample}} \\
DES13X1hav &	56551 &	-19.42 $\pm$ 0.32 &	2.6 &	29.62$^{+13.5}_{-6.65}$ & 11.10  \\
DES13X3gms &	56567 &	-19.66 $\pm$ 0.04 &	12.1 &	16.96$^{+1.95}_{-2.01}$ & 11.22  \\
DES14C3tvw &	57029 &	-19.69 $\pm$ 0.15 &	6.4 &	17.66$^{+4.69}_{-3.82}$ & 8.56  \\
DES14S2anq &	56903 &	-16.19 $\pm$ 0.02 &	8.6 &	20.50$^{+0.84}_{-0.85}$ & 10.44  \\
DES14S2plb &	56990 &	-15.76 $\pm$ 0.13 &	9.0 &	14.21$^{+101}_{-10.2}$ & 11.54  \\
DES14S2pli &	56987 &	-18.98 $\pm$ 0.04 &	5.2 &	29.83$^{+7.02}_{-5.3}$ & 11.84  \\
DES14X3pkl &	56987 &	-16.91 $\pm$ 0.10 &	5.4 &	14.84$^{+3.5}_{-3.65}$ & 7.47  \\
DES15C3lpq &	57317 &	-19.84 $\pm$ 0.08 &	9.2 &	16.62$^{+1.66}_{-1.44}$ & 9.54  \\
DES15C3mgq &	57331 &	-17.30 $\pm$ 0.04 &	3.3 &	11.90$^{+1.00}_{-0.90}$ & 5.30  \\
DES15E2nqh &	57366 &	-19.74 $\pm$ 0.06 &	6.5 &	16.30$^{+3.34}_{-2.44}$ & 11.57  \\
DES15S1fli &	57276 &	-19.76 $\pm$ 0.06 &	4.9 &	13.49$^{+5.77}_{-4.39}$ & 9.59  \\
DES15S1fll &	57276 &	-18.23 $\pm$ 0.05 &	5.8 &	13.43$^{+3.02}_{-2.35}$ & 12.71  \\
DES15X3mxf &	57356 &	-19.64 $\pm$ 0.02 &	7.7 &	9.98$^{+0.511}_{-0.472}$ & 6.84  \\
DES16C1cbd &	57660 &	-19.46 $\pm$ 0.05 &	9.7 &	19.38$^{+59.4}_{-13.5}$ & 10.85  \\
DES16C2ggt &	57756 &	-18.41 $\pm$ 0.05 &	3.8 &	16.57$^{+3.16}_{-3.76}$ & 7.31  \\
DES16E1bir &	57653 &	-22.24 $\pm$ 0.07 &	4.3 &	8.78$^{+3.58}_{-2.15}$ & 5.45  \\
DES16E2pv &	57629 &	-19.84 $\pm$ 0.08 &	3.5 &	26.81$^{+12.7}_{-6.96}$ & 9.76  \\
DES16S1dxu &	57717 &	-16.04 $\pm$ 0.08 &	6.2 &	24.94$^{+2.8}_{-2.77}$ & 11.16  \\
DES16X1eho &	57722 &	-20.97 $\pm$ 0.05 &	4.5 &	7.09$^{+1.87}_{-1.66}$ & 4.35  \\
DES16X3cxn &	57694 &	-19.62 $\pm$ 0.05 &	5.0 &	16.41$^{+4.89}_{-2.99}$ & 7.82  \\
\hline
\multicolumn{6}{c}{\textbf{Silver Sample}} \\
DES13C1tgd &	56615 &	-16.96 $\pm$ 0.27 &	6.7 &	- & - \\
DES13C3bcok &	56653 &	-19.01 $\pm$ 0.04 &	5.2 &	19.27$^{+1.3}_{-1.38}$ & 8.64  \\
DES13C3uig &	56625 &	-19.00 $\pm$ 0.08 &	4.9 &	8.98$^{+2.36}_{-1.93}$ & 5.18  \\
DES13E2lpk &	56575 &	-19.19 $\pm$ 0.07 &	5.4 &	17.29$^{+3.29}_{-3.33}$ & 8.02  \\
DES13X3npb &	56676 &	-19.01 $\pm$ 0.16 &	10.0 &	13.25$^{+4.66}_{-2.58}$ & 8.97  \\
DES13X3nyg &	56591 &	-20.26 $\pm$ 0.05 &	7.0 &	10.23$^{+2.39}_{-1.69}$ & 7.79  \\
DES13X3pby &	56602 &	-20.18 $\pm$ 0.07 &	3.6 &	- & -  \\
DES14X1bnh &	56915 &	-20.17 $\pm$ 0.16 &	6.5 &	16.55$^{+10.5}_{-5.04}$ & 8.19  \\
DES15C2eal &	57276 &	-17.30 $\pm$ 0.12 &	8.2 &	8.00$^{+3.73}_{-1.91}$ & 6.70  \\
DES15C3lzm &	57327 &	-18.54 $\pm$ 0.04 &	7.5 &	9.06$^{+0.67}_{-0.60}$ & 6.51  \\
DES15C3nat &	57356 &	-19.37 $\pm$ 0.14 &	5.9 &	13.24$^{+5.67}_{-3.67}$ & 6.98  \\
DES15C3opk &	57389 &	-20.20 $\pm$ 0.04 &	8.3 &	14.45$^{+2.98}_{-2.16}$ & 8.51  \\
DES15C3opp &	57385 &	-18.11 $\pm$ 0.15 &	6.2 &	9.63$^{+1.77}_{-1.45}$ & 6.32  \\
DES15X2ead &	57273 &	-17.25 $\pm$ 0.10 &	6.5 &	27.89$^{+9.57}_{-7.17}$ & 12.63  \\
DES16C3axz &	57643 &	-17.36 $\pm$ 0.10 &	4.9 &	12.72$^{+2.03}_{-2.09}$ & 6.24  \\
DES16C3gin &	57775 &	-19.15 $\pm$ 0.03 &	14.1 &	11.51$^{+2.14}_{-1.45}$ & 12.96  \\
DES16X3ega &	57730 &	-19.25 $\pm$ 0.03 &	12.6 &	18.17$^{+0.90}_{-0.92}$ & 14.91  \\

\hline
\end{tabular}
\end{adjustbox}
\label{tab:photometric}
\end{table}

\begin{figure}
\includegraphics[width=0.48\textwidth]{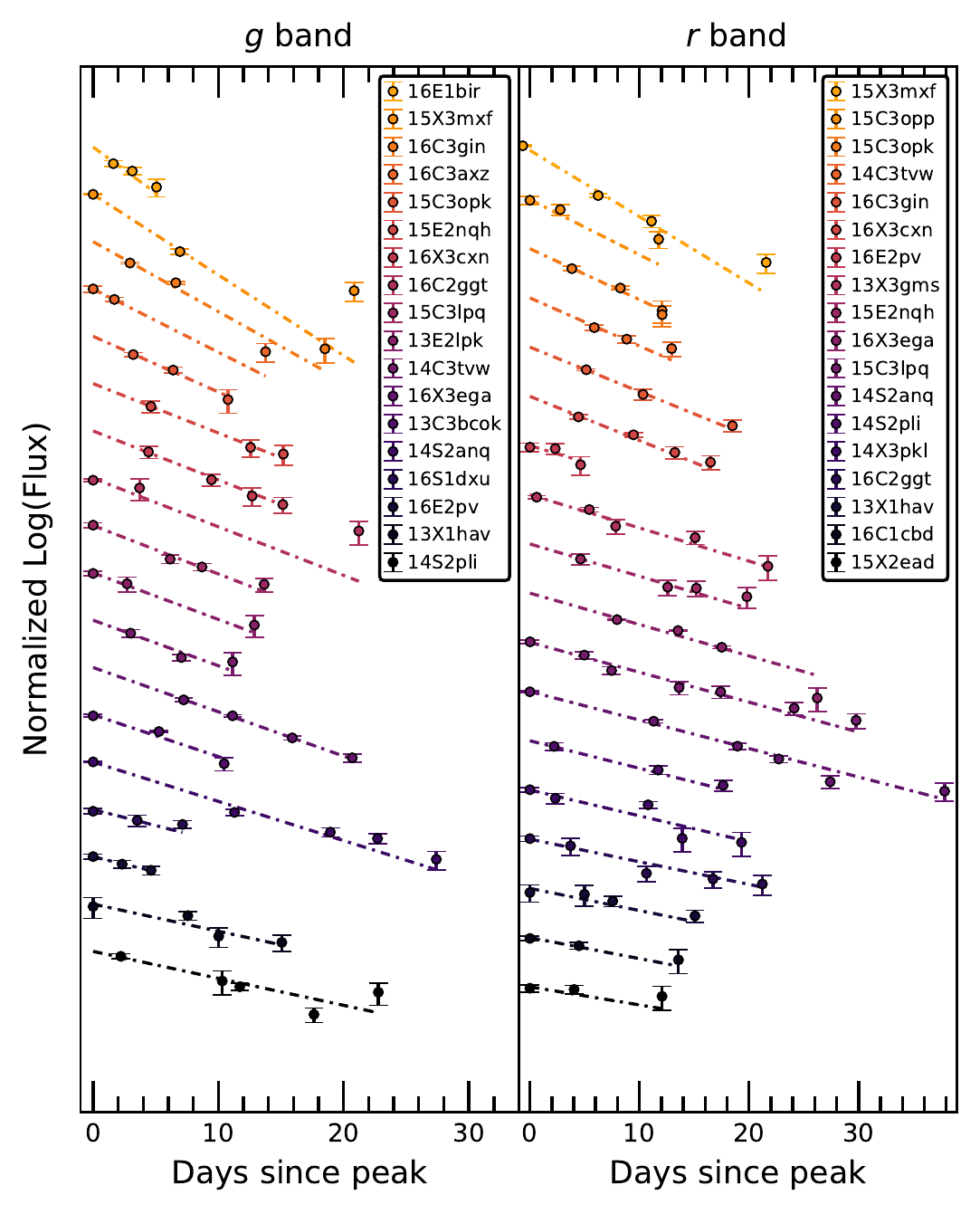}
\caption{Exponential fits to light curve declines. Only events for which more than two data points with S-to-N ratio $>3$ were used for the fit are shown. The best fits are plotted up till the last shown data point for each event.}
\label{fig:log_flux_t_decl}
\end{figure}

\begin{figure}
\includegraphics[width=0.48\textwidth]{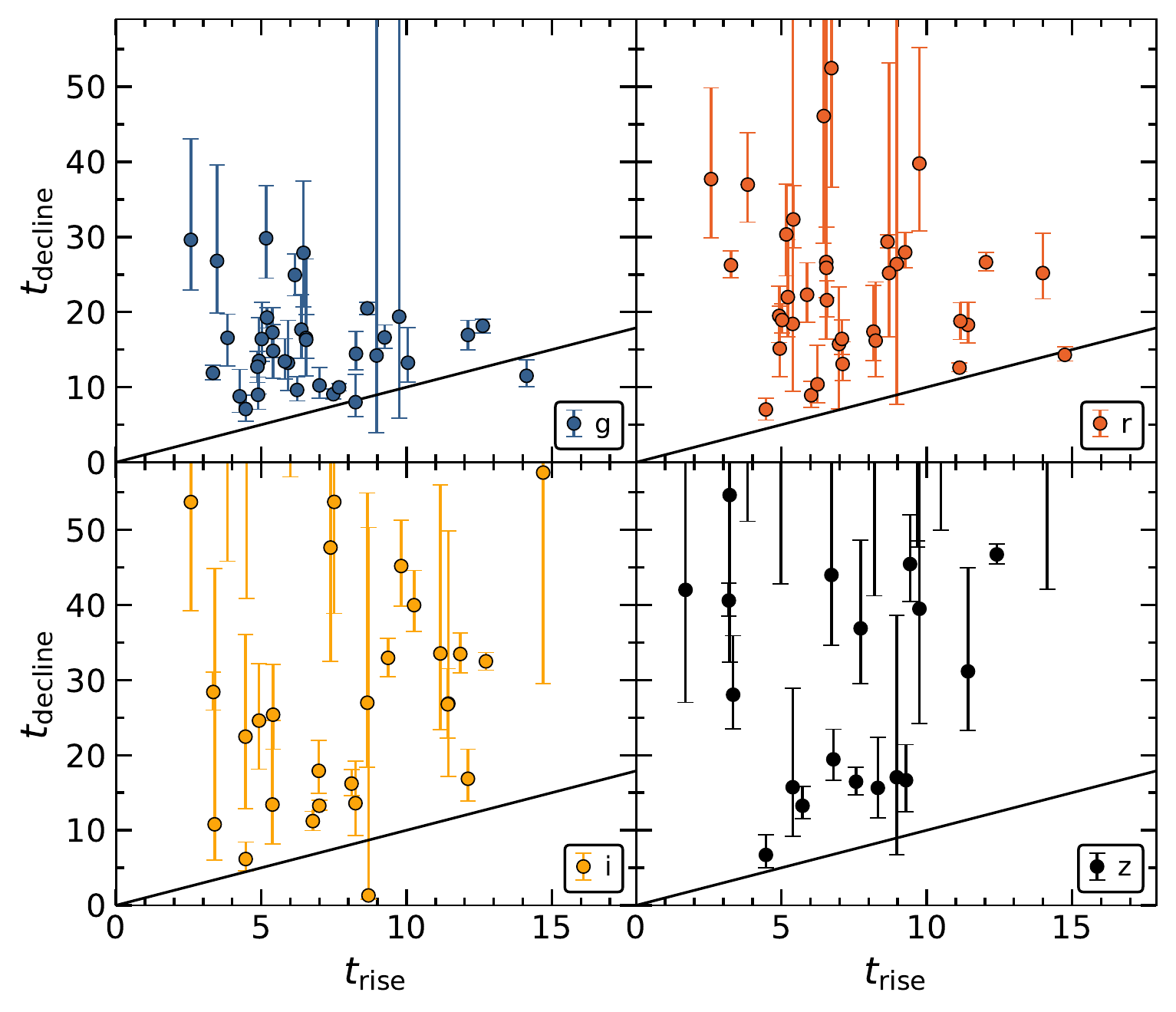}
\caption{$t_\mathrm{decline}$ against $t_\mathrm{rise}$ for gold and silver samples transients in all four bands. Black solid line represents $t_\mathrm{decline}=t_\mathrm{rise}$.}
\label{fig:t_dec_rise}
\end{figure}

\begin{figure}
\includegraphics[width=0.48\textwidth]{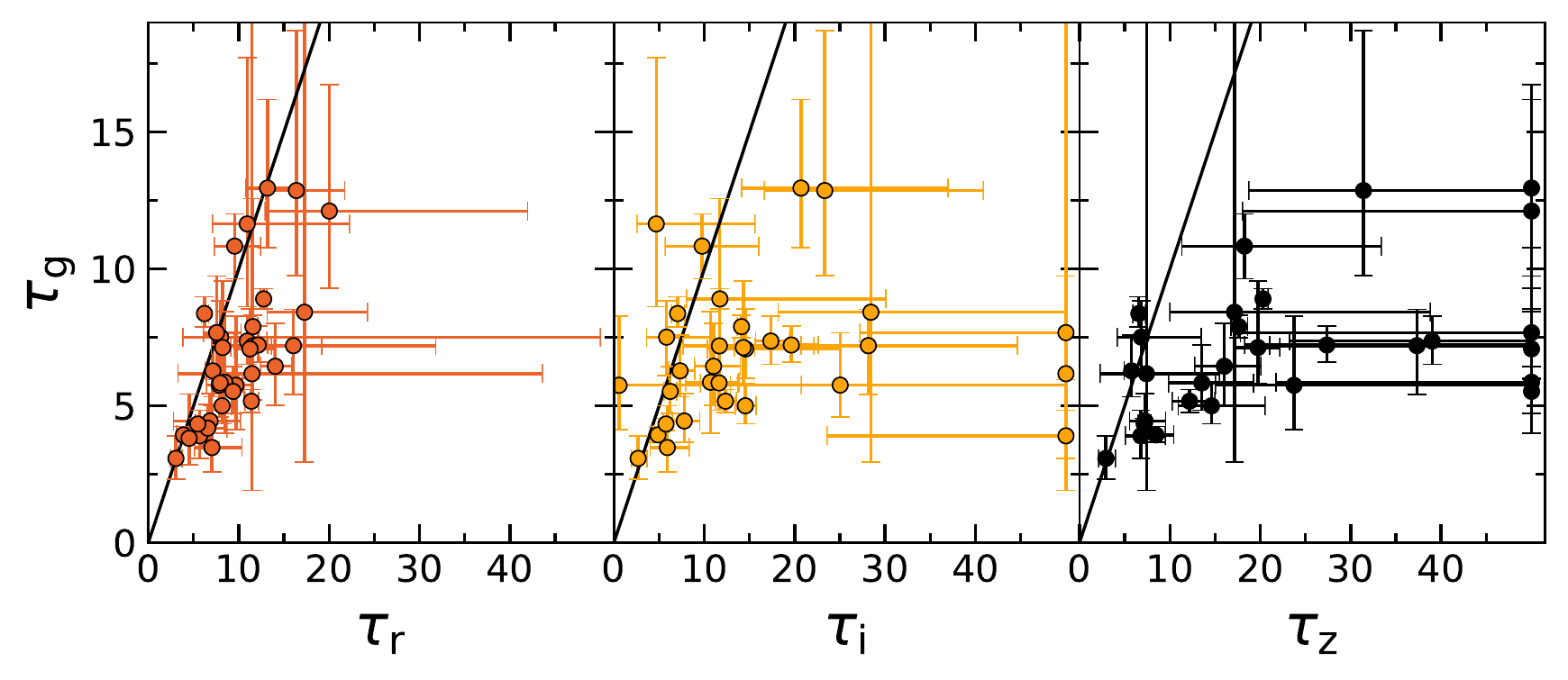}
\caption{Decay timescale $\tau_\mathrm{g}$ of gold and silvers samples plotted against $\tau_\mathrm{r}$, $\tau_\mathrm{i}$ and $\tau_\mathrm{z}$. Black solid line represents equal timescales.}
\label{fig:taus}
\end{figure}

We defined the rise time, $t_\mathrm{rise}$, to be the time between the observed peak and the last non-detection ($<3\sigma$) before the peak in each band in rest frame. We chose this definition because DES-SN has a week-long cadence and the majority of our transients were first detected at the apparent peak -- thus $t_\mathrm{rise}$ is likely an upper limit for the majority of the objects.   This long cadence also precludes adopting a method such as that employed by \citet{Drout2014}, who estimated the rise times for their sample by linearly interpolating the data points in the rise.  

We defined the decline time, $t_\mathrm{decline}$, as the time taken to decline to 1/10 of the observed peak flux for a best fitting exponential decline fit (which takes into account the first two non-detections after peak). The same fit can be parametrised by an exponential decay timescale $\tau$ \citep[see e.g.][]{Arnett1982}. The uncertainties for decline times were estimated with a Monte Carlo approach where we generated 1000 realizations of the light curves based on errors on our data and measured the spread of the distribution of fitted decline times across all realizations. Errors were then assumed to be the 16th and 84th percentile of the cumulative distribution function. In Figure \ref{fig:log_flux_t_decl} we have plotted exponential fits in the cases when more than two data points with S-to-N ratio $>3$ were used for the fits. In general fits used also the non-detections after the peak (fitting was done in linear flux space), but these are not shown in the Figure. As seen in the Figure, the exponential fits describe the data well.  Time above half maximum, $t_\mathrm{1/2}$, was then found based on the rise and decline times. Both rise and decline times are determined with respect to the observed maximum, and therefore are subject to the seven day cadence in the observer frame.

\begin{figure}
\includegraphics[width=0.48\textwidth]{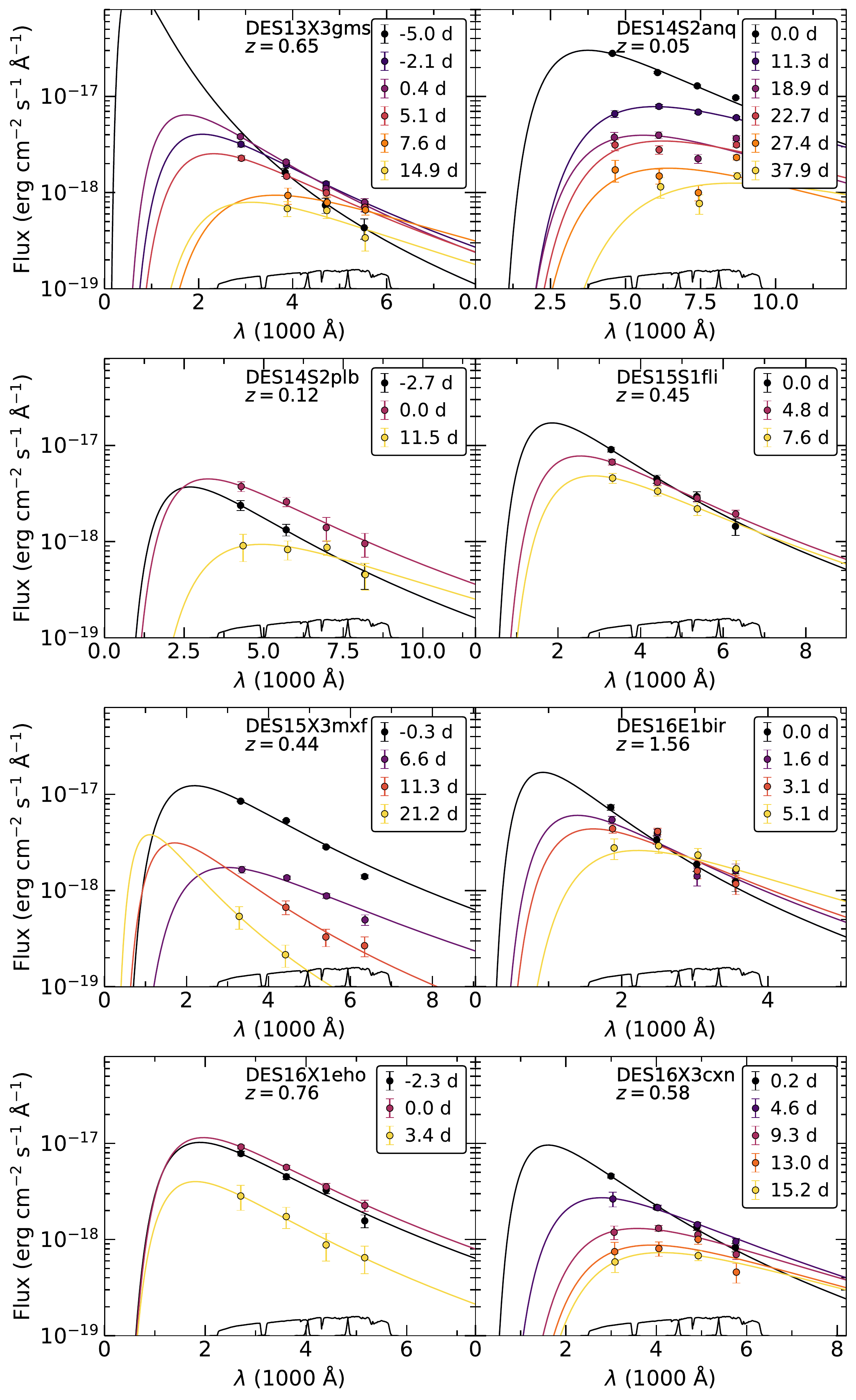}
\caption{Example blackbody fits to eight gold sample transients. Times of individual fits are given in rest frame days since the observed peak in $g$ band.  In general blackbody matches the data well around the peak, but at later times the best fits diverge from the data.}
\label{fig:BBs}
\end{figure}

Even though our estimates on rise times are likely upper limits, the majority of the transients still have $t_\mathrm{rise}\lesssim 10$ d. In fact, in the gold and silver samples only four objects have longer rise times. Out of these, DES13X3gms has a ~10 day gap in $g$-band observations before the peak, DES15X3mxf and DES16C3gin have a detection on the rise and thus the first non-detection is more than 10 days before peak.  However, we have very good photometric coverage of the rising light curve itself for one of our transients, and thus we can directly constrain the rise time. DES16X3ega has clear detections in $g$ and $i$ bands roughly 10 days before the peak and thus rise time is longer than that (see light curve in Figure \ref{fig:silver_lcs}).

In general the decline times are longer than the rise times, as can be seen in Figure~\ref{fig:t_dec_rise} where we plot the decline times against rise times in each band. Similar behavior was also reported by \citet{Drout2014}. The exponential decline timescales also appear to be longer the redder the band in question is, as seen in Figure~\ref{fig:taus} where we plot $\tau_\mathrm{r}$, $\tau_\mathrm{i}$ and $\tau_\mathrm{z}$ against $\tau_\mathrm{g}$. However, as these events are intrinsically fainter in redder bands constraining the decline for these bands is difficult.

As seen in Figure \ref{fig:t_dec_rise}, the decline timescales for our gold and silver sample transients span a wide range, for instance $\tau_\mathrm{g}$ is found to be between 3 and 15 days. Such a wide range is difficult to explain with any single value for the decline timescale. While $^{56}$Ni decay ($\tau_\mathrm{Ni}=8.764$), is roughly consistent for some of our events, it is not consistent for the whole sample. This suggests that the decay from nickel is not the primary source of the emission.

\subsection{Temperature and Radius Evolution}
\label{subsec:bb}

For each of the 37 gold and silver sample transients, we performed blackbody fits to the multi-band photometry for every 1.5 rest frame day ``epoch''  that had data in at least two bands. We fit for the temperature $T$ and radius $R$ of the transient at a given epoch using the fluxes observed in each band $F_\lambda$ to perform a $\chi^2$ minimization of a blackbody model which obeys the following equations:

\begin{ceqn}
\begin{gather}
\label{eq:bb}
F_\lambda(T, R) = \frac{2\pi h c^2}{\lambda_\mathrm{eff}^5} \cdot \frac{1}{e^{\frac{hc}{\lambda_\mathrm{eff} k T}}-1} \cdot \left(\frac{R}{D_\mathrm{l}}\right)^{2}, \\
\lambda_\mathrm{eff} =  \frac{\int_{\lambda_\mathrm{l}}^{\lambda_\mathrm{h}} t(\lambda) \cdot B(\frac{\lambda}{1+z}, T) \cdot \lambda \cdot d\lambda}{\int_{\lambda_\mathrm{l}}^{\lambda_\mathrm{h}} t(\lambda) \cdot B(\frac{\lambda}{1+z}, T) \cdot d\lambda}, \\
B(\lambda, T) =  \frac{2\pi h c^2}{\lambda^5} \cdot \frac{1}{e^{\frac{hc}{\lambda k T}}-1},
\end{gather}
\end{ceqn}
where $D_\mathrm{l}$ is the luminosity distance and $t(\lambda)$ is the transmission function over the band in question and $\lambda_\mathrm{l}$ and $\lambda_\mathrm{h}$ the wavelength limits for that band. Note that, the blackbody fits are done using the effective wavelength, $\lambda_\mathrm{eff}$, of each band. The effective wavelength depends not only on the redshift of the event, but also on the shape of blackbody emission. To estimate the shape, the corresponding blackbody has to be calculated in the rest frame ($\lambda/(1+z)$) and then shifted in to the observer frame. 

\begin{table}
\def\arraystretch{1.2}%
\setlength\tabcolsep{5pt}
\fontsize{8}{10}\selectfont
\caption{The best fitting parameters for a blackbody at peak. For DES13C3bcok best fitting temperature is found at the upper boundary for fitting. Last column (No.) refers to the number of data points the fit was based on.}
\centering
\begin{tabular}{l c c c c}
\hline
Name	&	$T_\mathrm{peak}$	&	$r_\mathrm{peak}$	&	$L_\mathrm{peak}$ &  No. \\
		&	(1000 K) &	($10^{14}$ cm) & ($10^{43}$ erg/s) & 		\\
\hline     
\multicolumn{4}{c}{\textbf{Gold Sample}} \\
DES13X1hav	& 12.64$^{+7.17}_{-3.07}$	& 10.89$^{+7.28}_{-4.93}$	& 2.16$^{+1.83}_{-0.35}$  & 3 \\
DES13X3gms	& 16.67$^{+1.22}_{-1.00}$	& 7.45$^{+0.66}_{-0.64}$	& 3.05$^{+0.33}_{-0.23}$  & 4 \\
DES14C3tvw	& 11.89$^{+8.95}_{-3.22}$	& 8.44$^{+5.74}_{-4.18}$	& 1.01$^{+1.5}_{-0.22}$   & 3\\
DES14S2anq	& 7.73$^{+0.17}_{-0.15}$	& 6.60$^{+0.22}_{-0.23}$	& 0.11$^{+0.01}_{-0.01}$  & 4 \\
DES14S2plb	& 8.91$^{+1.45}_{-1.02}$	& 4.12$^{+1.00}_{-0.87}$	& 0.08$^{+0.01}_{-0.01}$  & 4 \\
DES14S2pli	& 20.43$^{+3.96}_{-2.34}$	& 4.66$^{+0.68}_{-0.80}$	& 2.69$^{+1.12}_{-0.52}$  & 3 \\
DES14X3pkl	& 11.06$^{+1.29}_{-0.83}$	& 4.56$^{+0.57}_{-0.64}$	& 0.22$^{+0.03}_{-0.02}$  & 4 \\
DES15C3lpq	& 26.11$^{+6.06}_{-4.28}$	& 4.50$^{+0.90}_{-0.81}$	& 6.71$^{+3.80}_{-1.99}$  & 4 \\
DES15C3mgq	& 12.71$^{+0.78}_{-0.72}$	& 4.36$^{+0.37}_{-0.35}$	& 0.35$^{+0.03}_{-0.03}$  & 3 \\
DES15E2nqh	& 22.00$^{+4.12}_{-2.81}$	& 5.53$^{+1.03}_{-0.9}$		& 5.10$^{+1.91}_{-1.07}$  & 4 \\
DES15S1fli	& 15.62$^{+2.32}_{-1.52}$	& 9.17$^{+1.43}_{-1.46}$	& 3.56$^{+0.77}_{-0.44}$  & 4 \\
DES15S1fll	& 23.28$^{+6.97}_{-4.1}$	& 3.13$^{+0.69}_{-0.65}$	& 2.06$^{+1.57}_{-0.65}$  & 4 \\
DES15X3mxf	& 13.22$^{+0.31}_{-0.30}$	& 11.45$^{+0.44}_{-0.38}$	& 2.85$^{+0.08}_{-0.07}$  & 4 \\
DES16C1cbd	& 14.04$^{+1.1}_{-0.77}$	& 9.25$^{+0.91}_{-0.98}$	& 2.37$^{+0.21}_{-0.14}$  & 4 \\
DES16C2ggt	& 15.08$^{+1.76}_{-1.40}$	& 5.43$^{+0.77}_{-0.70}$	& 1.09$^{+0.21}_{-0.14}$  & 4 \\
DES16E1bir	& 32.17$^{+6.52}_{-4.17}$	& 7.36$^{+1.44}_{-1.39}$	& 41.41$^{+15.2}_{-7.72}$ & 4  \\
DES16E2pv	& 18.99$^{+2.81}_{-2.25}$	& 6.46$^{+1.28}_{-1.06}$	& 3.87$^{+0.99}_{-0.56}$  & 4 \\
DES16S1dxu	& 13.14$^{+2.40}_{-1.50}$	& 2.42$^{+0.45}_{-0.44}$	& 0.12$^{+0.04}_{-0.02}$  & 4 \\
DES16X1eho	& 14.76$^{+0.88}_{-0.79}$	& 16.52$^{+1.60}_{-1.39}$	& 9.24$^{+0.54}_{-0.45}$  & 4 \\
DES16X3cxn	& 18.01$^{+1.76}_{-1.34}$	& 6.63$^{+0.73}_{-0.72}$	& 3.29$^{+0.53}_{-0.34}$  & 4 \\
\hline
\multicolumn{4}{c}{\textbf{Silver Sample}} \\
DES13C1tgd	& 11.47$^{+7.70}_{-3.12}$	& 4.38$^{+2.72}_{-1.97}$	& 0.24$^{+0.35}_{-0.07}$  & 4 \\
DES13C3bcok	& 100.00$^{+0.00}_{-54.5}$	& 1.31$^{+0.93}_{-0.01}$	& 122.37$^{+3.67}_{-107}$ & 2  \\
DES13C3uig	& 20.46$^{+5.66}_{-2.97}$	& 4.03$^{+0.97}_{-0.97}$	& 2.02$^{+1.05}_{-0.40}$  & 3 \\
DES13E2lpk	& 12.42$^{+1.06}_{-0.95}$	& 10.24$^{+1.45}_{-1.18}$	& 1.78$^{+0.15}_{-0.13}$  & 4 \\
DES13X3npb	& 15.71$^{+5.17}_{-3.15}$	& 6.31$^{+2.51}_{-1.88}$	& 1.73$^{+1.05}_{-0.41}$  & 3 \\
DES13X3nyg	& 18.65$^{+1.83}_{-1.30}$	& 8.12$^{+0.84}_{-0.92}$	& 5.69$^{+0.84}_{-0.54}$  & 4 \\
DES13X3pby	&19.55$^{+2.03}_{-1.66}$	&6.90$^{+1.02}_{-0.88}$		& 4.95$^{+0.82}_{-0.49}$  & 4  \\
DES14X1bnh	&15.43$^{+3.76}_{-2.54}$	&10.44$^{+3.93}_{-2.99}$	& 4.39$^{+1.32}_{-0.56}$  & 3  \\
DES15C2eal	& 9.71$^{+1.16}_{-0.91}$	& 6.98$^{+1.3}_{-1.09}$		& 0.31$^{+0.04}_{-0.03}$  & 4 \\
DES15C3lzm	& 15.14$^{+1.00}_{-0.89}$	& 5.71$^{+0.46}_{-0.43}$	& 1.22$^{+0.14}_{-0.11}$  & 3 \\
DES15C3nat	& 31.74$^{+21.8}_{-8.48}$	& 2.60$^{+0.97}_{-0.92}$	& 4.88$^{+11.5}_{-2.25}$  & 4 \\
DES15C3opk	& 16.05$^{+1.21}_{-0.86}$	& 10.38$^{+0.83}_{-0.97}$	& 5.09$^{+0.54}_{-0.33}$  & 2 \\
DES15C3opp	& 28.42$^{+54.8}_{-10.3}$	& 2.07$^{+1.21}_{-1.13}$	& 1.99$^{+28.6}_{-1.15}$  & 2 \\
DES15X2ead	& 11.14$^{+1.29}_{-1.17}$	& 5.32$^{+1.03}_{-0.76}$	& 0.31$^{+0.05}_{-0.04}$  & 4 \\
DES16C3axz	& 36.01$^{+56.8}_{-12.6}$	& 1.21$^{+0.56}_{-0.57}$	& 1.76$^{+19.2}_{-1.10}$  & 2 \\
DES16C3gin	& 13.58$^{+0.52}_{-0.45}$	& 8.35$^{+0.39}_{-0.41}$	& 1.69$^{+0.10}_{-0.07}$  & 4 \\
DES16X3ega	& 16.41$^{+1.16}_{-0.93}$	& 7.32$^{+0.55}_{-0.55}$	& 2.77$^{+0.37}_{-0.25}$  & 3 \\
\hline
\end{tabular}
\label{tab:bb_peak}
\end{table}

\begin{figure}
\includegraphics[width=0.48\textwidth]{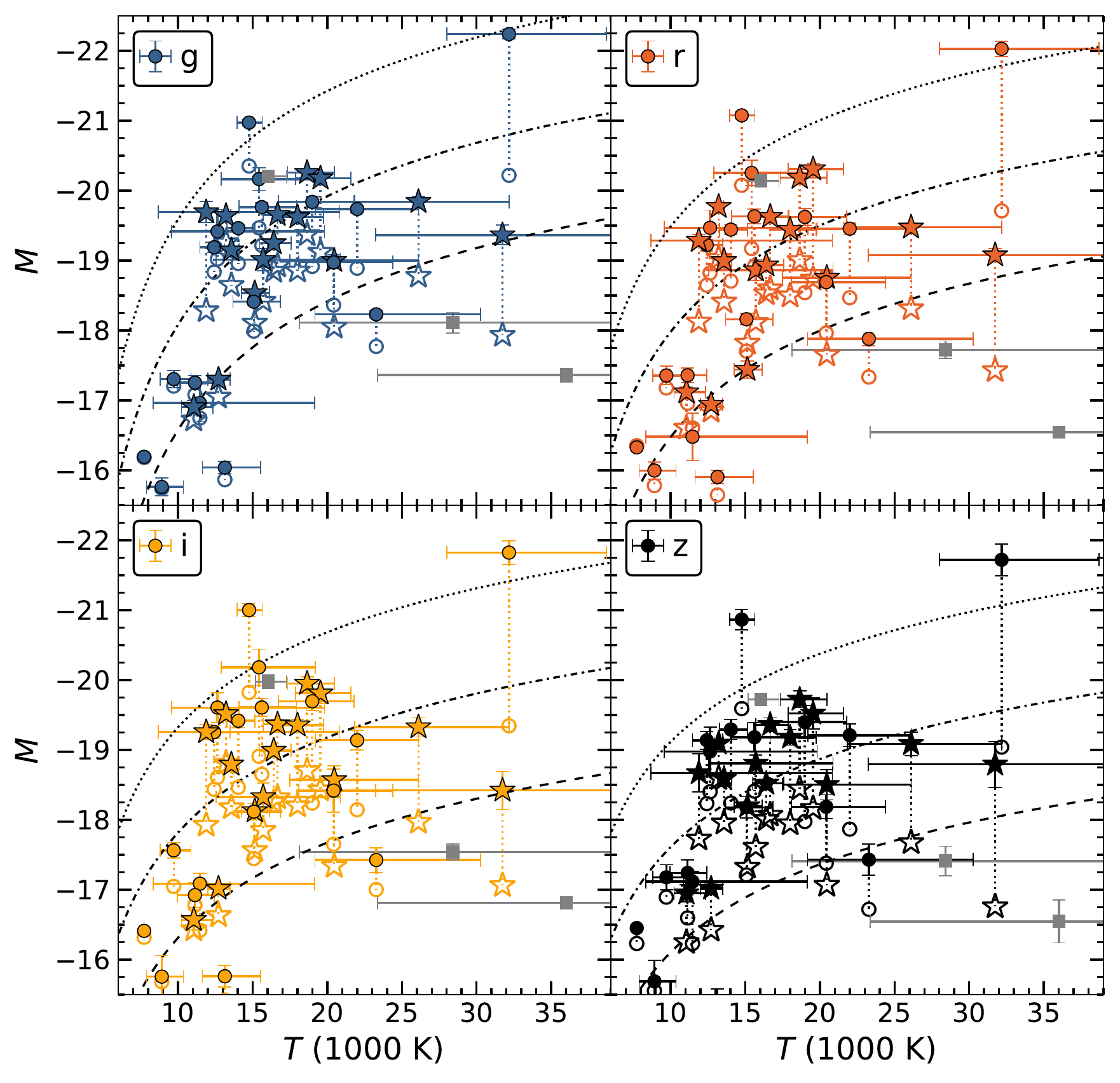}
\caption{Peak absolute magnitude against best fitting temperature at peak. Peak magnitudes are taken from the data points closest to the peak in $g$ band. Definition of markers is the same as in Figure \ref{fig:M_z}, with the addition of grey squares referring to events for which the blackbody fits at peak were based on only two bands. Note that DES13C3bcok is ignored in this plot due to high best fitting temperature. Errors for temperatures are quoted at 1$\sigma$ confidence. The three dashed lines correspond to absolute magnitudes in rest frame for constant radii of $r=5\cdot 10^{14}$ cm (bottom),  $r= 10^{15}$ cm and  $r=2\cdot 10^{15}$ cm (top).}
\label{fig:M_T} 
\end{figure}

\begin{figure*}
\includegraphics[width=0.99\textwidth]{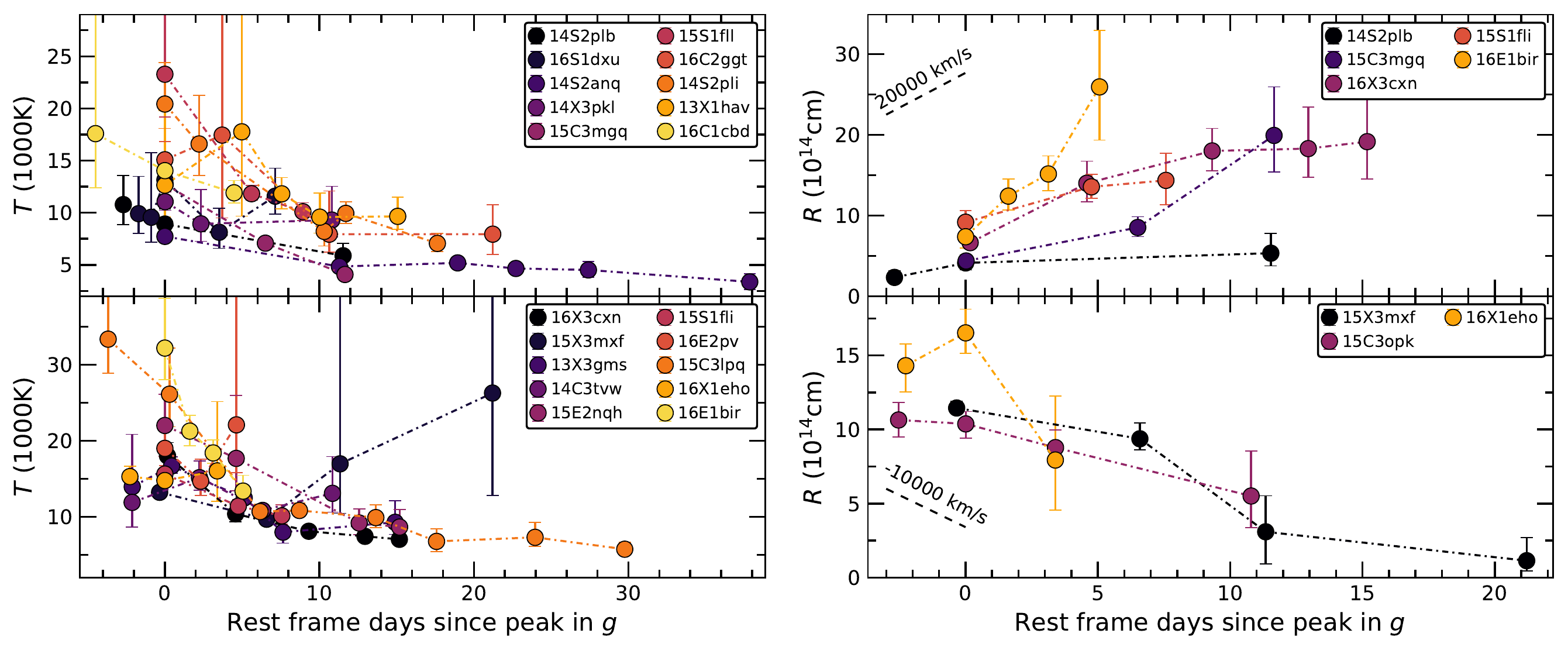}
\caption{On the left: Temperature evolution of gold sample transients, with events weaker than $M_\mathrm{g, peak}=-19.5$ plotted on top and brighter on bottom. On the right: Radius evolution for five example gold sample transients with increasing radii are plotted in the top panel and the three transients with decreasing radii are shown in the bottom panel. Note that DES15C3opk is a silver sample transient. Reference velocities are plotted with dashed lines. Brighter events are plotted sequentially in brighter colors in both panels.}
\label{fig:T_R}
\end{figure*}

\begin{figure}
\includegraphics[width=0.48\textwidth]{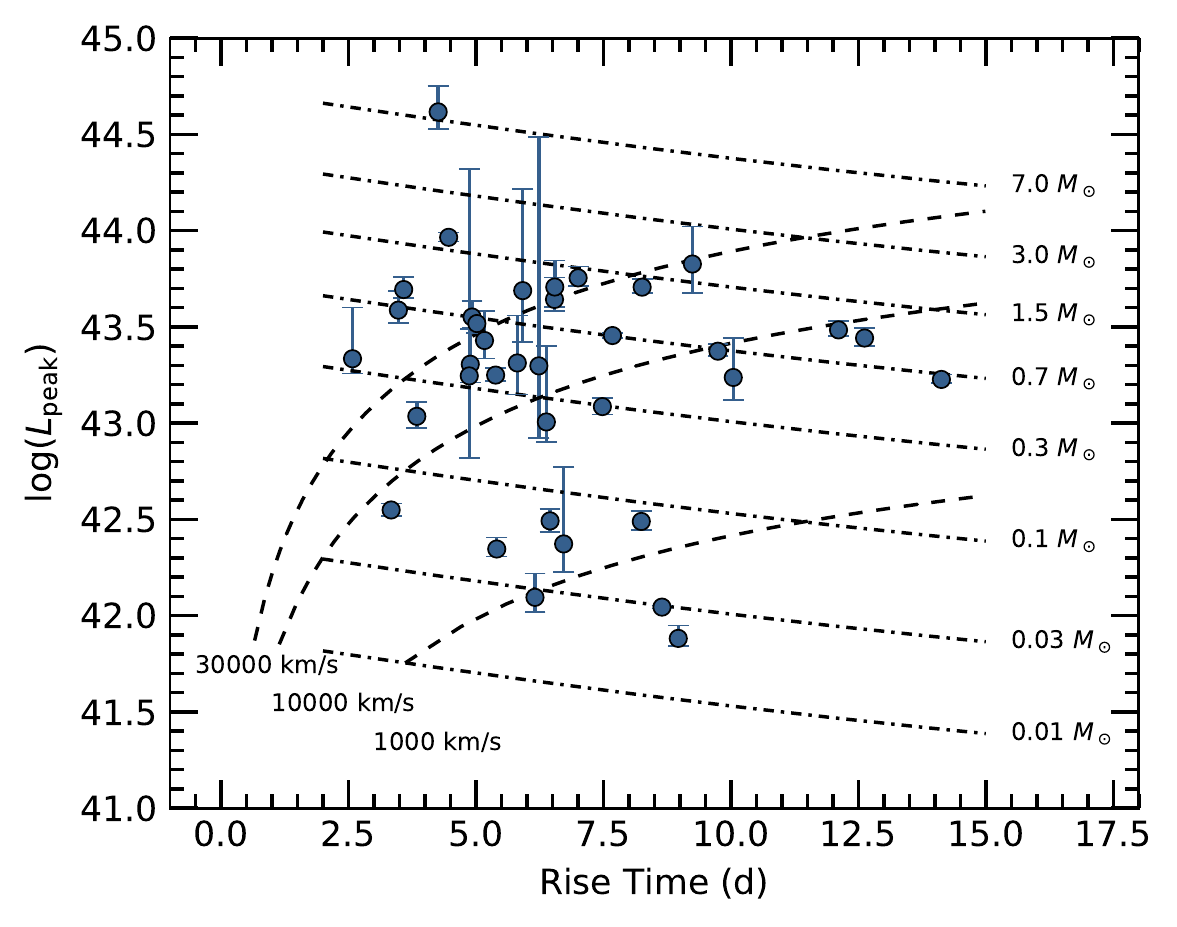}
\caption{Bolometric luminosity based on the blackbody fits (see Table. \ref{tab:bb_peak}) against rise time in $g$ band for gold and silver sample events. The nearly horizontal lines correspond to required nickel mass for given luminosity and rise time. The curved lines correspond to $M_\mathrm{Ejecta}=M_\mathrm{Ni}$ for a given expansion velocity assuming hydrogen and helium free ejecta ($\kappa=0.1$ cm$^2$ g$^{-1}$). All points above and left from given line are unrealistic for the given velocity.}
\label{fig:L_trise}
\end{figure}
This method returns temperature, $T$, and radius, $R$, for each epoch of observations. We then estimate $k$-corrected magnitudes by using the best-fit blackbody as the spectral energy distribution (SED) of the transient at that epoch. Additionally, this blackbody fit allows us to simply calculate the total bolometric luminosity at a given epoch.

In Figure \ref{fig:BBs} we show the best fitting blackbody curves for eight of our gold sample events. In general, a blackbody describes the data well around the peak. However, in the case of DES14S2anq the data is fitted almost perfectly up to $t=11.3$ d, but afterwards blackbody provides a poor fit. Similarly the fits to  DES13X3gms and DES16X3cxn around two weeks after peak seem to be worse than at earlier times. Thus it is possible that the late phase emission of the rapid events is not well described by a blackbody.  In Table \ref{tab:bb_peak} we have given the best-fit blackbody temperatures and radii at peak with corresponding bolometric luminosities and number of data points fits were based on for our gold and silver sample events. The given uncertainties have been estimated with a Monte Carlo approach with 500 realizations.  

We find temperatures ranging from 8000 K to 30000 K, and radii between a few $10^{14}$ cm and a few tens of $10^{14}$ cm. These values are mostly consistent with the ones found in literature. \citet{Arcavi2016} and \citet{Whitesides2017} discussed temperature and radius evolution for bright objects ($M_\mathrm{g}\approx -20$), finding these to be cooler (with temperatures around $10000$ K) and mostly larger (with $R\gtrsim 10^{15}$ cm) than transients in our sample with comparable absolute magnitudes. On the other hand, \citet{Drout2014} found the temperatures to be slightly higher when compared with objects of similar brightness in our sample, and therefore the found radii are several times smaller. Nonetheless, the peak bolometric luminosities estimated based on the blackbody fits are found in the range $\sim 10^{42}-10^{44}$ erg/s and are comparable with other rapidly evolving transients. 

The peak temperature correlates well with the absolute magnitude, as illustrated in Figure~\ref{fig:M_T}. The majority of our events are clustered around 15000 K, while the faintest objects have temperatures below $10000$ K and the brightest are found at $\sim20000-30000$ K.  The brightness of any blackbody is dependent on both its temperature and radius. To examine the impact of radius on the brightness of our objects, we plot curves of constant radius in the figure. These curves are in rest frame and thus correspond to the magnitudes before $k$-corrections, which are plotted as open markers. The blackbody fits at peak for DES15C3opk, DES15C3opp and DES16C3axz were based on two bands -- thus we plot these transients as grey squares without $k$-corrected values. 

In the left column of Figure~\ref{fig:T_R}, we plot the temperature evolution of the gold sample transients (split into two plots for visual clarity).  From these curves we can clearly see that the temperatures decrease in time. For many objects, the drop in temperature is accompanied by an increase in radius, as shown in the top right panel of Figure~\ref{fig:T_R} for five gold sample events. The rate of change of the photosphere based on these curves imply velocities ranging from $v\lesssim10000$ km/s for DES14S2plb up to $v\approx40000$ km/s for DES16E1bir. These values are not the true velocities of the ejecta, due to the changing opacity of the expanding material, but can be used as a lower limit for the expansion. Such high velocities constrain the explosion time to be $\lesssim 5$ d before the peak in several cases. For instance, the radius of DES16E1bir recedes to zero in roughly two days, when assuming the expansion rates to be constant. Similar high velocities have been reported for other rapidly evolving transients, for example iPTF16asu expanded at $34500 \pm 5400$ km/s \citep{Whitesides2017}, but also for broad-lined SNe Ic (SNe Ic-bl) with velocities up to 30000 km/s \citep[see e.g.][]{Modjaz2011}.

However, the radius does not increase for all of our events. In the bottom right panel of Figure~\ref{fig:T_R} we show three transients for which the radius appears to be decreasing with time while the temperature either stays constant or slightly increases. One of these events is DES16X1eho for which the radius decreases from $R\approx16\cdot 10^{14}$ cm at peak to $R\approx8\cdot 10^{14}$ cm in just 3.4 days after the peak.  The decrease in radius could be explained by a shock breakout in a optically thick circumstellar wind surrounding a star. Such a scenario will be discussed in Section \ref{sec:discussions}. 

Using models from \citet{Arnett1982} and \citet{Stritzinger2005}, the peak bolometric luminosity and the rise time of any transients can be used to investigate if nickel decay is the primary power source for the emission. Under this scenario, in order to have higher peak brightness, more nickel needs to be produced during the explosion. However, if the ejecta mass is increased, the diffusion timescale through the ejecta, and hence the rise time of the light curve, increase as well. This is a problem for most of our transients. To demonstrate this, we have plotted the peak luminosity (based on the blackbody fits, see Table \ref{tab:bb_peak}) against the rise time for our gold and silver sample transients values in Figure \ref{fig:L_trise} with reference lines assuming constant nickel mass from $0.01M_\odot$ to $7M_\odot$. We also plotted lines for $M_\mathrm{Ejecta}=M_\mathrm{Ni}$ with given velocities assuming hydrogen and helium free ejecta ($\kappa=0.1$ cm$^2$ g$^{-1}$). All points above and left of a given curve is unrealistic as $M_\mathrm{Ejecta}<M_\mathrm{Ni}$. In general, the nickel decay can produce the peak luminosities for most of our events, but the ejecta would have to be both rapidly expanding and composed almost completely of nickel. The nickel decay could potentially be the power source for our faintest events, if the total nickel mass is very low. However, we want to emphasize that our rise time estimates are effectively upper limits (see Section \ref{subsec:t_rise_dec}) and therefore the scenario is unlikely. For the reference lines we have assumed spherical symmetry, optically thick ejecta, nickel distribution peaked at the center of the ejecta and constant density profile of the ejecta
\begin{figure*}
\includegraphics[width=\textwidth]{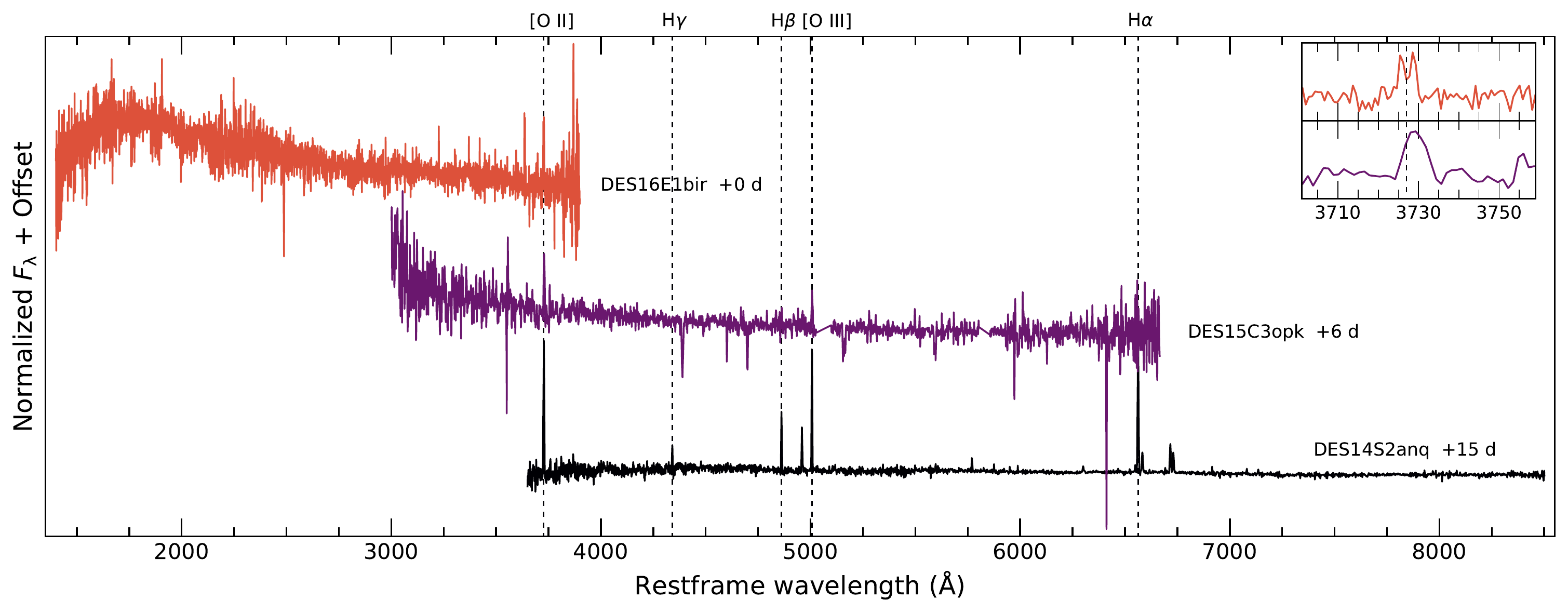}
\caption{Spectra for DES14S2anq ($z=0.05$), DES15C3opk ($z=0.57$) and DES16E1bir ($z=1.56$). Phases are given in rest frame days since the peak in $g$  band. Spectrum for DES14S2anq is dominated by host galaxy features, while spectra for DES15C3opk and DES16E1bir have strong underlying blue continuum. Inset: [O II] $\lambda$3727 Å doublet line for DES16E1bir and DES15C3opk.}
\label{fig:spectra}
\end{figure*}

\begin{figure*}
\includegraphics[width=\textwidth]{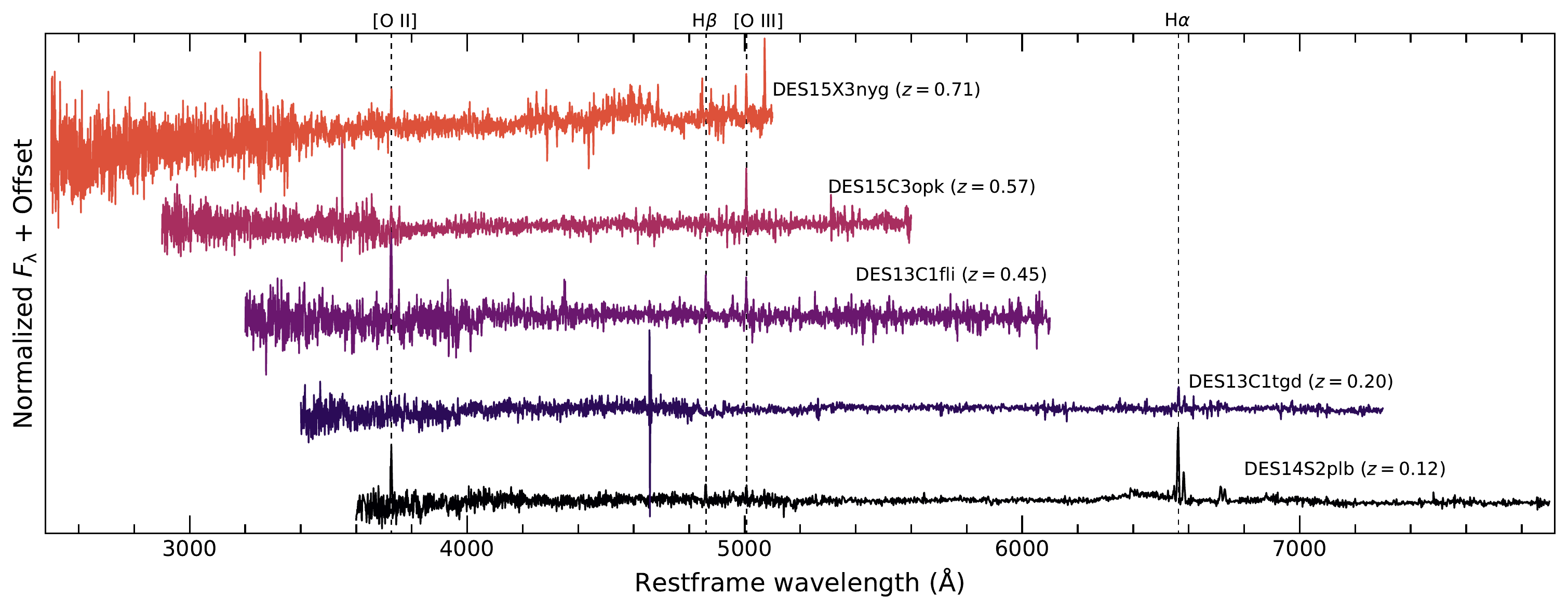}
\caption{Host galaxy spectra for five gold and silver sample transients.}
\label{fig:host_spectra}
\end{figure*}

\section{Spectroscopic Properties}
\label{sec:spectra}

Spectra of DES14S2anq, DES15C3opk, and DES16E1bir are plotted in Figure~\ref{fig:spectra}.  The spectra of DES15C3opk (at +6 d after peak) and DES16E1bir (at peak) are dominated by underlying blue continuum with only few lines present.  The visible lines in the spectrum of DES15C3opk are [O II] $\lambda$3727 Å doublet  and [O III] $\lambda$5007 Å lines with a trace of H$\beta$, all coinciding with $z=0.57$. The observed lines arise most likely from the host galaxy (see Sec. \ref{sec:hostgal}). The [O II] doublet line is also present in the spectrum of DES16E1bir, where this line has been used to estimate the redshift $z=1.56$ (see inset of Figure \ref{fig:spectra}). In the Figure we also present the spectrum for DES14S2anq, but this spectrum was obtained two weeks after the peak and is thus dominated by host galaxy features.

The strong blue continuum is a typical feature in the spectra of previously discovered rapidly evolving transients at least around peak \citep[see e.g.][]{Arcavi2016, Whitesides2017}. Given that all of our events appear to be very hot, and hence produce strong ionizing emission, we would expect to see some narrow permitted lines if the transient is surrounded by a low-density optically thin region. The lack of such lines may provide important clues to the configuration of the progenitor system. Such a featureless continuum has also been observed in the early spectra of CCSNe. \citet{Khazov2016} analyzed early spectra ($\lesssim10$ d after explosion) of 84 type II SNe and found that 27 of them had blue featureless spectra and 12 had high ionization lines with an underlying blue continuum. These lines included mostly Balmer and He II lines as typical for type II, but also O II and O III lines in a few occasions. They concluded these lines were created by recombination of some circumstellar medium (CSM) flash-ionized by a shock breakout.

\section{Host Galaxy Properties}
\label{sec:hostgal}

\begin{figure}
\includegraphics[width=0.48\textwidth]{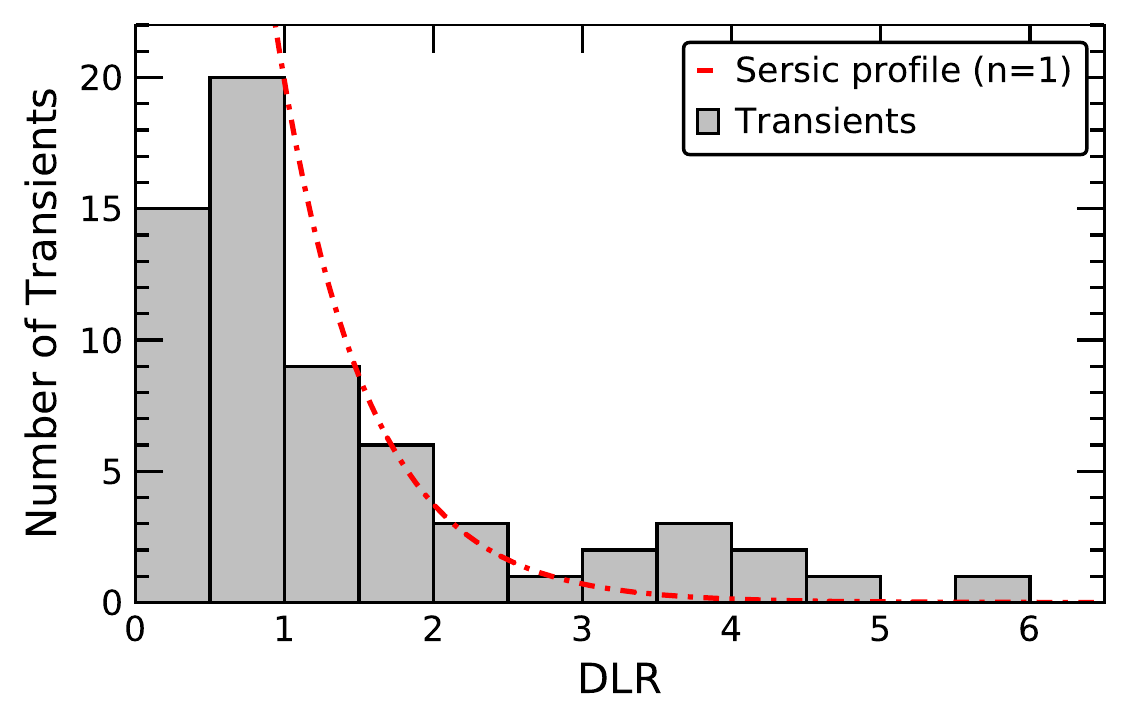}
\caption{DLR distribution of the 63 events with host galaxies. Sersic profile for $n=1$ corresponding to spiral galaxies has been plotted in red dashed line. }
\label{fig:dlr_hist}
\end{figure}

In this section we discuss the host galaxy properties for all 63 events that appear to be associated with a galaxy. We have readily available OzDES spectra for the host galaxies of 27 of the 37 gold+silver sample transients, and we plot some representative host spectra in Figure~\ref{fig:host_spectra}. The remaining 10 spectroscopic redshifts are from various different surveys as shown in Table 3. As most of them have released only the redshifts, we have decided to focus on the spectra from OzDES. Based on the 27 host spectra, all events appear to originate in star forming galaxies, as indicated by the presence of nebular emission lines. 

We are confident in the host identification for our transients, based on the Directional Light Radius \citep[DLR,][]{Sullivan2006} distribution plotted in Figure~\ref{fig:dlr_hist} (see Table \ref{tab:z_sources}). This distribution matches a Sersic profile \citep{Sersic1963} for spiral galaxies ($n=1$) well between $\mathrm{DLR}\approx1-3$, providing additional evidence that these transients arise from star-forming host galaxies. Objects near their host centres ($\mathrm{DLR}\lesssim1$) may be under-represented, the host galaxy emission is very strong and therefore the weaker rapid events will likely be missed \citep[i.e., the classical ``Shaw Effect",][]{Shaw1979}.  Events distant from their hosts ($\mathrm{DLR}\gtrsim3$) are undoubtedly rare, but with no underlying host galaxy contamination we expect our discovery completeness to be very high. 

\begin{figure}
\includegraphics[width=0.48\textwidth]{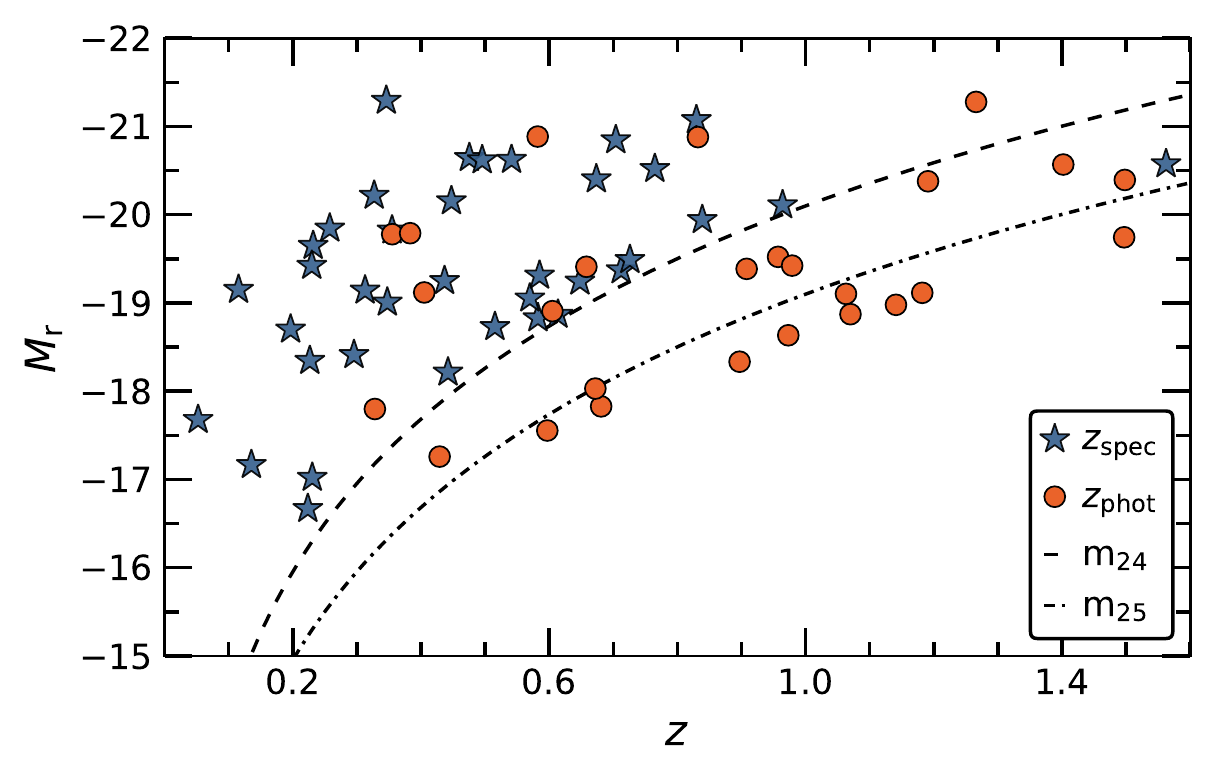}
\caption{Absolute magnitude of the host galaxy in $r$ band plotted against the redshift for the 63 transients with a host galaxy. Redshifts are spectroscopic for the hosts of the gold and silver sample events (star) and photometric for the hosts of the bronze sample events (circle). Dashed line corresponds to $m_{r}=24$ and dot dashed line to $m_{r}=25$.}
\label{fig:M_z_host}
\end{figure}

We calculate rough estimates of the stellar masses for our host galaxy sample using the DES-SN $griz$ photometry fitted with the code {\tt ZPEG} \citep{zpeg}.  This code is designed for estimating photometric redshifts, but performs galaxy SED template matching and thus simultaneously derives stellar masses.  It also provides a crude estimate of the galaxy's star-formation intensity, namely the specific star-formation rate (sSFR) based on that of the best fitting template. Spectroscopic redshifts have been used for fitting for the gold and silver sample events.  From these results presented in Table \ref{tab:host_gal_mass_sfr}, we see the significant majority of our sample are found in strongly star-forming galaxies ($\mathrm{sSFR} > -10.5$).  This implies a short-lived progenitor system.

\begin{table}
\def\arraystretch{1.3}%
\setlength\tabcolsep{2pt}
\fontsize{9}{11}\selectfont
\caption{Host galaxy log stellar masses ($M_*$ -- in units of $M_\odot$) and specific star formation rates (sSFR -- in units of yr$^{-1}$) of the best-fitting model.  Hosts with blank sSFR entries were best fit to a passive galaxy template.}
\begin{adjustbox}{max width=0.49\textwidth, center=0.45\textwidth}
\centering
\begin{tabular}{l c c p{0.5em} l c c}
\cline{1-3} \cline{5-7}
Name	&	log$(M_*/M_\odot)$	&	log(sSFR)	& & Name	&	log$(M_*/M_\odot)$ 	&	log(sSFR) \\
\cline{1-3} \cline{5-7}    
\multicolumn{7}{c}{\textbf{Gold Sample}} \\
DES13X1hav	&	$9.15_{-0.19}^{+0.02}$	&	$-9.32$	& &	DES15S1fli	&	$9.94_{-0.10}^{+0.02}$	&	$-9.38$\\
DES13X3gms	&	$9.23_{-0.05}^{+0.09}$	&	$-9.17$	& &	DES15S1fll	&	$9.27_{-0.09}^{+0.09}$	&	$-9.44$\\
DES14C3tvw	&	$11.15_{-0.06}^{+0.03}$	&	$-9.73$	& &	DES15X3mxf	&	$9.33_{-0.10}^{+0.02}$	&	$-9.38$\\
DES14S2anq	&	$9.00_{-0.01}^{+0.01}$	&	$-$		& &	DES16C1cbd	&	$11.21_{-0.07}^{+0.03}$	&	$-$\\
DES14S2plb	&	$9.92_{-0.11}^{+0.05}$	&	$-9.54$	& &	DES16C2ggt	&	$9.59_{-0.01}^{+0.01}$	&	$-9.50$\\
DES14S2pli	&	$9.82_{-0.28}^{+0.07}$	&	$-9.44$	& &	DES16E1bir	&	$8.26_{-0.38}^{+1.23}$	&	$-9.08$\\
DES14X3pkl	&	$9.33_{-0.01}^{+0.12}$	&	$-9.56$	& &	DES16E2pv	&	$9.79_{-0.02}^{+0.15}$	&	$-9.39$\\
DES15C3lpq	&	$9.23_{-0.01}^{+0.01}$	&	$-9.32$	& &	DES16S1dxu	&	$8.69_{-0.09}^{+0.02}$	&	$-9.44$\\
DES15C3mgq	&	$8.38_{-0.11}^{+0.18}$	&	$-9.38$	& &	DES16X1eho	&	$9.96_{-0.51}^{+0.14}$	&	$-9.25$\\
DES15E2nqh	&	$8.96_{-0.29}^{+0.33}$	&	$-9.32$ & &	DES16X3cxn	&	$9.35_{-0.12}^{+0.03}$	&	$-9.32$\\
\cline{1-3} \cline{5-7}  
\multicolumn{7}{c}{\textbf{Silver Sample}} \\
DES13C1tgd	&	$10.20_{-0.01}^{+0.01}$	&	$-10.18$	&& DES15C3lzm	&	$9.85_{-0.01}^{+0.01}$	&	$-9.44$ \\
DES13C3bcok	&	$10.79_{-0.32}^{+0.01}$	&	$-9.44$		&& DES15C3nat	&	$9.85_{-0.13}^{+0.26}$	&	$-9.34$ \\
DES13C3uig	&	$10.64_{-0.25}^{+0.05}$	&	$-9.59$		&& DES15C3opk	&	$9.84_{-0.01}^{+0.01}$	&	$-9.47$ \\
DES13E2lpk	&	$10.17_{-0.30}^{+0.07}$	&	$-9.38$		&& DES15C3opp	&	$8.96_{-0.10}^{+0.02}$	&	$-9.38$ \\
DES13X3npb	&	$10.67_{-0.02}^{+0.21}$	&	$-9.52$		&& DES15X2ead	&	$9.82_{-0.11}^{+0.03}$	&	$-9.50$ \\
DES13X3nyg	&	$9.06_{-0.06}^{+0.05}$	&	$-9.08$		&& DES16C3axz	&	$9.62_{-0.01}^{+0.09}$	&	$-9.50$ \\
DES13X3pby	&	$9.35_{-0.52}^{+0.09}$	&	$-9.17$		&& DES16C3gin	&	$9.56_{-0.09}^{+0.03}$	&	$-9.44$ \\
DES14X1bnh	&	$10.74_{-0.39}^{+0.36}$	&	$-9.37$		&& DES16X3ega	&	$10.02_{-0.01}^{+0.09}$	&	$-9.44$ \\
DES15C2eal	&	$8.41_{-0.10}^{+0.01}$	&	$-9.44$ && \\
\cline{1-3} \cline{5-7}  
\multicolumn{7}{c}{\textbf{Bronze Sample}} \\
DES13C1acmt	&	$9.27_{-0.35}^{+0.09}$	&	$-9.32$	&&	DES15E2lmq	&	$9.81_{-0.73}^{+0.14}$	&	$-9.38$	\\
DES13C3abtt	&	$10.00_{-0.39}^{+0.11}$	&	$-9.32$	&&	DES15X3atd	&	$9.85_{-0.01}^{+0.13}$	&	$-9.24$	\\
DES13C3asvu	&	$9.66_{-0.12}^{+0.19}$	&	$-9.08$	&&	DES15X3kyt	&	$9.28_{-1.82}^{+0.47}$	&	$-9.08$	\\
DES13C3avkj	&	$9.27_{-0.06}^{+0.11}$	&	$-9.08$	&&	DES15X3nlv	&	$9.14_{-0.02}^{+0.12}$	&	$-9.24$	\\
DES13C3nxi	&	$8.53_{-0.11}^{+0.03}$	&	$-9.32$	&&	DES15X3oma	&	$8.33_{-0.15}^{+0.34}$	&	$-9.08$	\\
DES13C3smn	&	$7.43_{-0.11}^{+2.21}$	&	$-9.00$	&&	DES16C1bde	&	$8.95_{-0.24}^{+0.07}$	&	$-9.17$	\\
DES13X2oyb	&	$9.98_{-0.38}^{+0.23}$	&	$-9.08$	&&	DES16C2grk	&	$10.00_{-0.23}^{+0.06}$	&	$-9.44$	\\
DES13X3aakf	&	$10.01_{-0.35}^{+0.10}$	&	$-9.38$	&&	DES16C3auv	&	$9.27_{-0.23}^{+0.03}$	&	$-9.17$	\\
DES13X3alnb	&	$8.53_{-0.33}^{+0.44}$	&	$-9.25$	&&	DES16C3cdd	&	$9.51_{-0.33}^{+0.29}$	&	$-9.03$	\\
DES14C1jnd	&	$8.52_{-0.43}^{+0.58}$	&	$-9.44$	&&	DES16X1ddm	&	$10.37_{-0.19}^{+0.02}$	&	$-9.77$	\\
DES14E2xsm	&	$9.62_{-0.14}^{+0.01}$	&	$-9.41$	&&	DES16X2bke	&	$10.37_{-0.37}^{+0.13}$	&	$-$	\\
DES15C3edw	&	$8.71_{-0.01}^{+0.16}$	&	$-9.32$	&&	DES16X3ddi	&	$10.04_{-0.27}^{+0.30}$	&	$-9.37$	\\
DES15C3pbi	&	$8.69_{-0.29}^{+0.32}$	&	$-8.90$	&&	DES16X3erw	&	$9.59_{-0.01}^{+0.01}$	&	$-9.51$	\\

\cline{1-3} \cline{5-7}
\end{tabular}
\end{adjustbox}
\label{tab:host_gal_mass_sfr}
\end{table}

In Figure~\ref{fig:M_z_host} we present the host galaxy absolute $r$-band magnitudes against redshift for all 63 events with either spectroscopic (star symbol) or photometric (circle) redshift. We can see that there is a clear cutoff in the number of objects with spectroscopic redshift above $z\approx 0.8$. This is due to the fact that we have galaxy redshifts only to events that occur in apparently bright galaxies ($m_\mathrm{r}\lesssim24$).  The reason why most of our bronze sample hosts lack $z_\mathrm{spec}$ is because they are fainter than that. Additionally, the efficiency of OzDES for getting host galaxy redshifts drops significantly above $z\sim0.8$ as key emission features are redshifted into wavelength ranges with strong night sky emission. Based on the photometric redshifts, 13/35 bronze sample events seem to be associated with a galaxy at $z_\mathrm{phot}\geq0.9$. In general, the photometric redshifts for our gold and silver samples hosts are consistent with the spectroscopic ones (see Table \ref{tab:z_sources}). Therefore, it is likely that at least some of these 13 hosts are found at high redshifts, making the brighter end of these transients more common than in our current spectroscopic sample.

\section{Discussions}
\label{sec:discussions}

Our analysis of 37 rapidly evolving transients with accurate distances shows that the observed properties span a wide range in both luminosity and timescales. Despite the apparent similarity of the light curve evolution, the objects are found to have very different brightnesses, temperatures, radii and expansion velocities. Thus if we assume these events to be similar in origin, the power source has to be flexible enough to explain this broad range of observed behaviours.

One physical scenario which has been proposed to explain the rapidly evolving light curves is a shock breakout and the consequent shock cooling in surrounding wind \citep[see e.g.][]{Ofek2010,Balberg2011,Chevalier2011,Ginzburg2013}.  In this scenario, a shock launched by the core collapse of a massive star travels through an extended layer of circumstellar material (CSM). At least some small amount of emission arising from cooling of shock-heated matter (in either the stellar envelope itself or some CSM) should be present in all SNe \citep{Nakar2010}, but the timescale and total emitted luminosity depend strongly on the progenitor structure. While the peak luminosity is dictated by the size and temperature of the event, the duration depends most strongly on the cooling mass \citep[see e.g.][]{Nakar2010,Rabinak2011}. 

The scenario of shock cooling in an extended material (e.g. extended envelope or wind) gives a natural explanation for the very early phases of the light curves for several double-peaked CCSNe such as the Type IIb SN 1993J \citep{Wheeler1993} which exhibits a rapid pre-peak seen in both blue and red bands \citep{Nakar2014}. This rapid pre-peak is then followed by a typical SN light curve powered by the decay of $^{56}$Ni. Each of these pre-peaks have brightness up to $M\approx-18$ in the bluer bands but last altogether only for $\lesssim 10$ d, unlike most of our events. For instance pre-peak of SN 1993J had $M_\mathrm{B}\approx-16.5$ and lasted for $\approx 5$ d and pre-peak of another Type IIb SN 2011dh was as bright as the main peak of the SN at $M_\mathrm{g}\approx-16.5$ and lasted for $\lesssim5$ d \citep[see e.g.][]{Arcavi2011}. However, even though such timescales are comparable to some of our events, no rapidly evolving transients in literature has been classified as Type IIb SNe to date. Similar rapid pre-peaks have also been discovered in SLSNe such as SN 2006oz \citep{Leloudas2012}, LSQ14bdq \citep{Nicholl2015} and DES14X3taz \citep{Smith2016}, where they are bright up to $M_\mathrm{g}\approx-20$. 

Shock breakout and the consequent shock cooling has been considered as a possible scenario for several previously discovered rapidly evolving transients \citep[see e.g.][]{Ofek2010, Drout2014, Arcavi2016}, but there are a few problems to be solved. \citet{Whitesides2017} argued that the light curve of iPTF 16asu can not be explained solely by a shock cooling model for extended material from \citet{Piro2015}. This model produces a short and fairly symmetric light curve, so iPTF16asu would therefore require emission from nickel decay in order to reproduce the light curve tail. A typical CCSN has a peak brightness of order $M_\mathrm{B}\lesssim -18$ \citep[see e.g.][]{Richardson2014} and can not be seen above $z\approx0.6$ in DES. Therefore, under this scenario our distant events should have only the short lived emission from shock cooling, without an exponential decline. This also means we cannot rule out an associated CCSN event for our high redshift rapid transients. Some of our distant events follow this expectation of short-lived emission (e.g. DES16X1eho and DES16E1bir), but several events at $z\approx0.6-0.7$ have clearly longer decline than rise times (e.g. DES14C3tvw, DES15C3lpq and DES15C3opk). These events would require exceptionally bright CCSNe to explain the declining light curve. 

Our sample also includes several faint nearby events that have smooth exponential declines without clear signatures of two separate power sources (see e.g. DES14S2anq and DES16X3ega in Appendix \ref{app:gold_lc}). Additionally, two of our closest events, DES14S2anq and DES14S2plb, have peak brightnesses of order that of a relatively faint CCSN ($M_\mathrm{r}\approx-16$). If nickel decay was the source of emission in the tail of these events (i.e. the peak of the the nickel decay light curve would be after the peak produced by shock breakout and shock cooling) the amount of $^{56}$Ni should be as small or smaller than in a faint CCSN to produce the brightness of the observed declining light curves ($M_\mathrm{r} \gtrsim-15$). Similarly, these events lack the common broad light curve -- FWHM $> 30$~days, caused by photon diffusion through several solar masses of ejecta -- of typical nickel-powered CCSNe. Thus if these events are typical CCSNe, they would have to all have very low $^{56}$Ni masses and ejecta masses.

Whether these rapid light curves might frequently accompany CCSN light curves remains uncertain. This analysis is biased against finding such events -- our characterization of events by a single light curve width was useful for finding rapidly evolving events, but might have caused us to miss events with double peaks.  This will be a topic of study for future analyses of the DES-SN transient sample.

In the light of the recent discovery of the neutron star (NS) merger GW170817 \citep[see e.g.][]{Alexander2017,Andreoni2017, Arcavi2017, Blanchard2017,Chornock2017, Coulter2017, Cowperthwaite2017, Drout2017, Fong2017,Kasliwal2017,Kilpatrick2017,Margutti2017,Nicholl2017, Siebert2017,Smartt2017,Soares-Santos2017}, it is interesting to compare its optical properties with our events. The kilonova has almost immediate rise to peak and a very rapid declining light curve, with last detections in $g$ band roughly a week after the merger \citep{Cowperthwaite2017,Drout2017,Soares-Santos2017}. The merger is also fairly faint with absolute magnitude $\approx-15.5$ in the optical bands. The SED is well described by a blackbody for the first few days, with a temperature $T\approx11000$ K and radius $R\approx3\cdot10^{14}$ cm at 0.5 days \citep{Kasliwal2017},  $T\approx8300$ K and $R\approx4.5\cdot10^{14}$ cm at 0.6 days \citep{Cowperthwaite2017} and $T=5500 \pm 150$ K and $R\approx7\cdot10^{14}$ cm at 1.5 days after the merger \citep{Nicholl2017}. Additionally \citet{Drout2017} found temperatures around 2500 K from 5.5 to 8.5 days after the merger. Consistent values are also given by \citet{Smartt2017}, who found that temperature decreased from $T=7600\pm2000$ K at 0.6 days to $T=1900\pm500$ K at 13.3 days.  In our gold and silver samples there are only two events which have comparable brightnesses with comparable values for the temperature and the radius at peak: DES14S2anq ($M_\mathrm{g}\approx-16.2$, $T\approx7700$ K, $R\approx6.6\cdot10^{14}$ cm) and DES14S2plb ($M_\mathrm{g}\approx-15.8$, $T\approx8900$ K, $R\approx4.1\cdot10^{14}$ cm). However, while the kilonova declines $\approx4.5$ mags in $g$ band and $\approx3.5$ mags in $r$ band in 5 days after the merger \citep[see e.g.][]{Drout2017}, the DES events decline only 1.2-1.5 in $g$ and 0.7-0.9 mags in $r$ in roughly 10 days after the peak brightness. Due to the significantly slower decline rate, it is unlikely that these events are associated with a NS-NS merger. Similarly, \citet{Siebert2017} compared the light curve of GW170817 to other optical transients, including several rapidly evolving events from \citet{Drout2014}, and concluded that the kilonova evolves much faster in comparison. On the other hand, the events in our bronze sample lack spectroscopic redshift as their host galaxies are mostly weaker than $m_\mathrm{r}\approx24$. In order to observe NS-NS merger, such as GW170817 with a peak magnitude on the order of $-15.5$, it would have to be at very low redshift. However, it is unlikely that these host galaxies would be at such a low redshifts without us having a spectroscopic redshift of them (unless their luminosities are significantly below the faintest typical galaxies and the photometric redshifts are catastrophically wrong). Therefore, it is unlikely that our bronze sample events are associated with a NS-NS merger. Additionally, \citet{Doctor2017} search the first two years of DES-SN data and found no kilonovae consistent with their model predictions, while \citet{scolnic17} used the observed kilonova light curve of GW170817 to calculate DES-SN should find 0.26 kilonovae in its 5-year lifetime. Such a low number is inconsistent with the total number of objects we discovered.

The large diversity in the peak luminosities we observe for our sample of rapid transients, despite their having similar light curves in observer frame, might be difficult to explain with a single physical process. Therefore, it is possible that there are several different subclasses with different physical mechanisms powering the emission. Such an analysis would be best facilitated by an even larger sample of rapidly evolving transients to distinguish the possible sub-classes from each other.  We plan to undertake such an analysis when distances (i.e., redshifts) are secured for the full DES-SN sample.                                                                                                            

\section{Conclusions}
\label{sec:conclusions}
We have presented a sample of  72 rapidly evolving transients found in the Dark Energy Survey. A subsample of 37 transients have a spectroscopic redshift from host galaxy features. This increases the total number of optical transients with $t_{1/2}\lesssim12$ d in rest frame by more than a factor of two making our sample the most substantial to date. 

The discovered events have a wide range of brightnesses ($-15.75>M_g>-22.25$) and distances ($0.05>z>1.56$). However, we note that due to their fast rise times, it is likely that our observational epochs did not exactly coincide with the true epoch of peak brightness, and we therefore expect that they are actually slightly brighter at true peak.  The light curves are characterized by a very rapid rise in $\lesssim10$ d and by a slightly longer exponential decline. The decline timescales also appear to be longer in the redder bands and do not reconcile well with the decay timescale of $^{56}$Ni.

Photometric data of the events is well fitted with a blackbody up to roughly two weeks after the peak. The photospheres appear to be cooling and expanding rapidly in time, starting from high temperatures (up to $30000$ K) and large radii (up to $10^{15}$ cm). The absolute brightness also correlates well with the peak temperature, with the faintest events having peak temperatures below $10000$ K and brightest above $20000$ K. However, a few transients appear to have a receding photosphere while the temperature stays roughly constant (DES15X3mxf, DES15C3opk and DES16X1eho). The bolometric peak luminosities based on the blackbody fits are in the range $\sim10^{42}-10^{44}$ erg/s. The spectra at peak (and slightly after) are dominated by a featureless blue continuum with some host galaxy emission lines present -- this spectral behaviour is to be expected in the case of a hot, optically thick medium. 

All of our events with detected host galaxies are found in star forming galaxies, implying a short-lived progenitor system. Therefore, the currently favored scenario to explain these rapid events is a shock breakout and consequent shock cooling in optically thick, low mass circumstellar wind surrounding a CCSN. Shock cooling in extended material has been used to explain pre-peaks in several Type II SNe \citep[e.g. SN 1993J][]{Wheeler1993} and SLSNe \citep[e.g. DES14X3taz]{Smith2016}. The breakout and consequent cooling has already been considered for rapidly evolving events by \citet{Ofek2010} and \citet{Drout2014}. Moreover, the observed featureless blue spectra of these rapid transients is consistent with early spectra of transients powered by shock breakout cooling in either envelope or CSM. The decreasing photospheric radii could also be explained with this scenario. If the wind was originally just dense enough to be optically thick, the rapid expansion could drop the density so drastically that the observed radii would actually decrease. However, the current models fail to produce the exponential decline and might require additional energy source to power the light curve tail \citep{Whitesides2017}. Additionally, we want to emphasize that due to wide range of rise and decline timescales and peak luminosities it is difficult to describe all the events with a single model. Therefore it is likely that the events in our sample are produced by more than one physical mechanism.

In future analyses, we will explore the light curve modeling of these peculiar events. This will include both testing and comparing the existing shock breakout light curve models, but this will require an examination of the subtleties introduced by the limitations of the models themselves \citep[see e.g.][]{Rabinak2011, Piro2015}, including the limited treatment of radiative transfer effects.  We are also aiming to understand the luminosity function better, for which we need both new events with known redshifts, as well as spectroscopic redshifts for the hosts of our bronze sample events. Knowing the total brightness range of rapid events would help us to understand the physical limitations needed for the powering engine. We are also planning to explore the possibility of having two energy sources: one powering the bright peak and another the exponential decay. We will search the DES-SN data for any events that appear to have either two peaks with rapid timescales or have a clear cut off in a light curve after peak. As DES-SN has recently concluded its fifth observing year, we are also hoping to obtain both early and late time spectrum to shed light on these peculiar events.

\section*{Acknowledgements}

We acknowledge support from EU/FP7-ERC grant no [615929]. Based in part on data obtained from the ESO Science Archive Facility under program 097.D-0709. Parts of this research were conducted by the Australian Research Council Centre of Excellence for All-sky Astrophysics (CAASTRO), through project number CE110001020.

Funding for the DES Projects has been provided by the U.S. Department of Energy, the U.S. National Science Foundation, the Ministry of Science and Education of Spain, 
the Science and Technology Facilities Council of the United Kingdom, the Higher Education Funding Council for England, the National Center for Supercomputing 
Applications at the University of Illinois at Urbana-Champaign, the Kavli Institute of Cosmological Physics at the University of Chicago, 
the Center for Cosmology and Astro-Particle Physics at the Ohio State University,
the Mitchell Institute for Fundamental Physics and Astronomy at Texas A\&M University, Financiadora de Estudos e Projetos, 
Funda{\c c}{\~a}o Carlos Chagas Filho de Amparo {\`a} Pesquisa do Estado do Rio de Janeiro, Conselho Nacional de Desenvolvimento Cient{\'i}fico e Tecnol{\'o}gico and 
the Minist{\'e}rio da Ci{\^e}ncia, Tecnologia e Inova{\c c}{\~a}o, the Deutsche Forschungsgemeinschaft and the Collaborating Institutions in the Dark Energy Survey. 

The Collaborating Institutions are Argonne National Laboratory, the University of California at Santa Cruz, the University of Cambridge, Centro de Investigaciones Energ{\'e}ticas, 
Medioambientales y Tecnol{\'o}gicas-Madrid, the University of Chicago, University College London, the DES-Brazil Consortium, the University of Edinburgh, 
the Eidgen{\"o}ssische Technische Hochschule (ETH) Z{\"u}rich, 
Fermi National Accelerator Laboratory, the University of Illinois at Urbana-Champaign, the Institut de Ci{\`e}ncies de l'Espai (IEEC/CSIC), 
the Institut de F{\'i}sica d'Altes Energies, Lawrence Berkeley National Laboratory, the Ludwig-Maximilians Universit{\"a}t M{\"u}nchen and the associated Excellence Cluster Universe, 
the University of Michigan, the National Optical Astronomy Observatory, the University of Nottingham, The Ohio State University, the University of Pennsylvania, the University of Portsmouth, 
SLAC National Accelerator Laboratory, Stanford University, the University of Sussex, Texas A\&M University, and the OzDES Membership Consortium.

Based in part on observations at Cerro Tololo Inter-American Observatory, National Optical Astronomy Observatory, which is operated by the Association of 
Universities for Research in Astronomy (AURA) under a cooperative agreement with the National Science Foundation.

The DES data management system is supported by the National Science Foundation under Grant Numbers AST-1138766 and AST-1536171.
The DES participants from Spanish institutions are partially supported by MINECO under grants AYA2015-71825, ESP2015-66861, FPA2015-68048, SEV-2016-0588, SEV-2016-0597, and MDM-2015-0509, 
some of which include ERDF funds from the European Union. IFAE is partially funded by the CERCA program of the Generalitat de Catalunya.
Research leading to these results has received funding from the European Research
Council under the European Union's Seventh Framework Program (FP7/2007-2013) including ERC grant agreements 240672, 291329, and 306478.
We  acknowledge support from the Australian Research Council Centre of Excellence for All-sky Astrophysics (CAASTRO), through project number CE110001020, and the Brazilian Instituto Nacional de Ci\^encia
e Tecnologia (INCT) e-Universe (CNPq grant 465376/2014-2).
MJC acknowledges the support of the UK Science and Technology Facilities Council (STFC) under grant (ST/P000398/1).

This manuscript has been authored by Fermi Research Alliance, LLC under Contract No. DE-AC02-07CH11359 with the U.S. Department of Energy, Office of Science, Office of High Energy Physics. The United States Government retains and the publisher, by accepting the article for publication, acknowledges that the United States Government retains a non-exclusive, paid-up, irrevocable, world-wide license to publish or reproduce the published form of this manuscript, or allow others to do so, for United States Government purposes.




\bibliographystyle{mnras}
\bibliography{bib} 
\clearpage




\appendix
\section{Rest Frame Light Curves of Gold and Silver Sample Transients}
\label{app:gold_lc}

\begin{minipage}{\textwidth}
\centering
\includegraphics[width=0.98\textwidth]{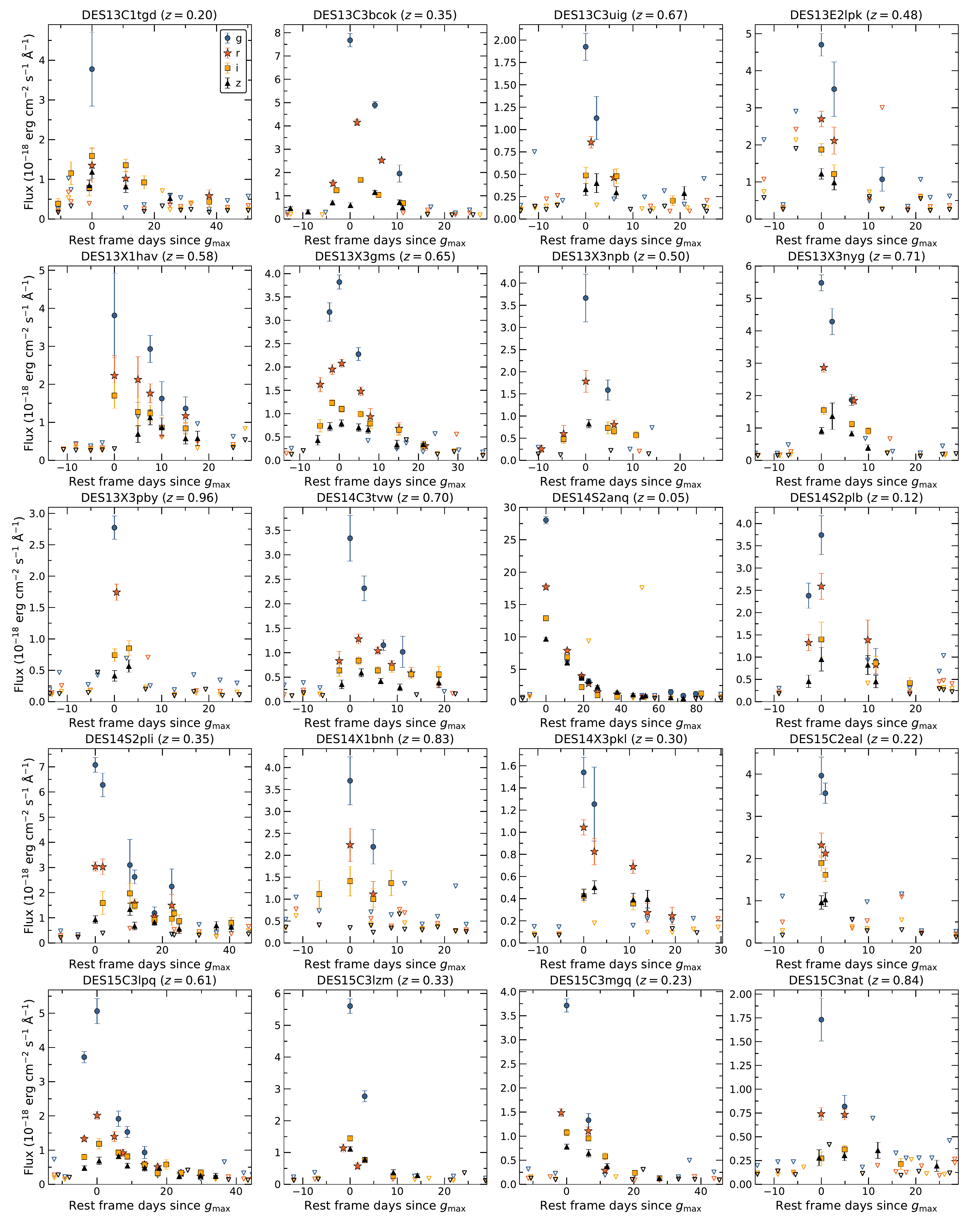}
\captionsetup{labelfont=bf, font=small, justification=centering, labelsep=period}
\captionof{figure}{Rest frame light curves of gold and silver sample transients. Open triangles represent 1$\sigma$ error of data points below 3$\sigma$ detection. }
\label{fig:gold_lcs_all}
\end{minipage}

\clearpage

\addtocounter{figure}{-1}

\begin{minipage}{\textwidth}
\includegraphics[width=0.98\textwidth]{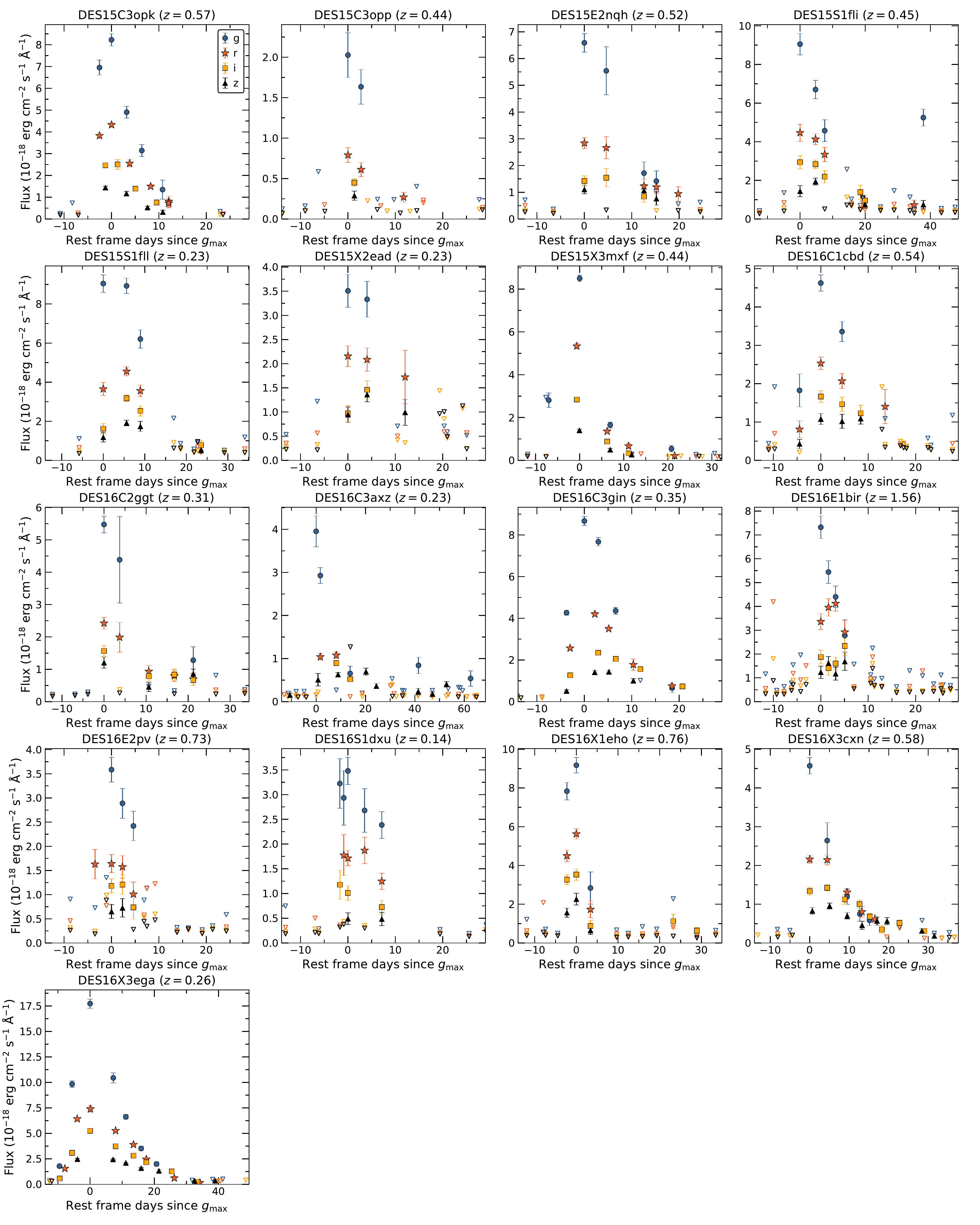}
\captionsetup{labelfont=bf, font=small, justification=centering, labelsep=period}
\captionof{figure}{(Continued.)}
\label{fig:gold_lcs_all}
\end{minipage}

\clearpage

\section{Observer Frame Light Curves of Bronze Sample Transients}
\label{app:bronze_lc}

\begin{minipage}{\textwidth}
\centering
\includegraphics[width=0.98\textwidth]{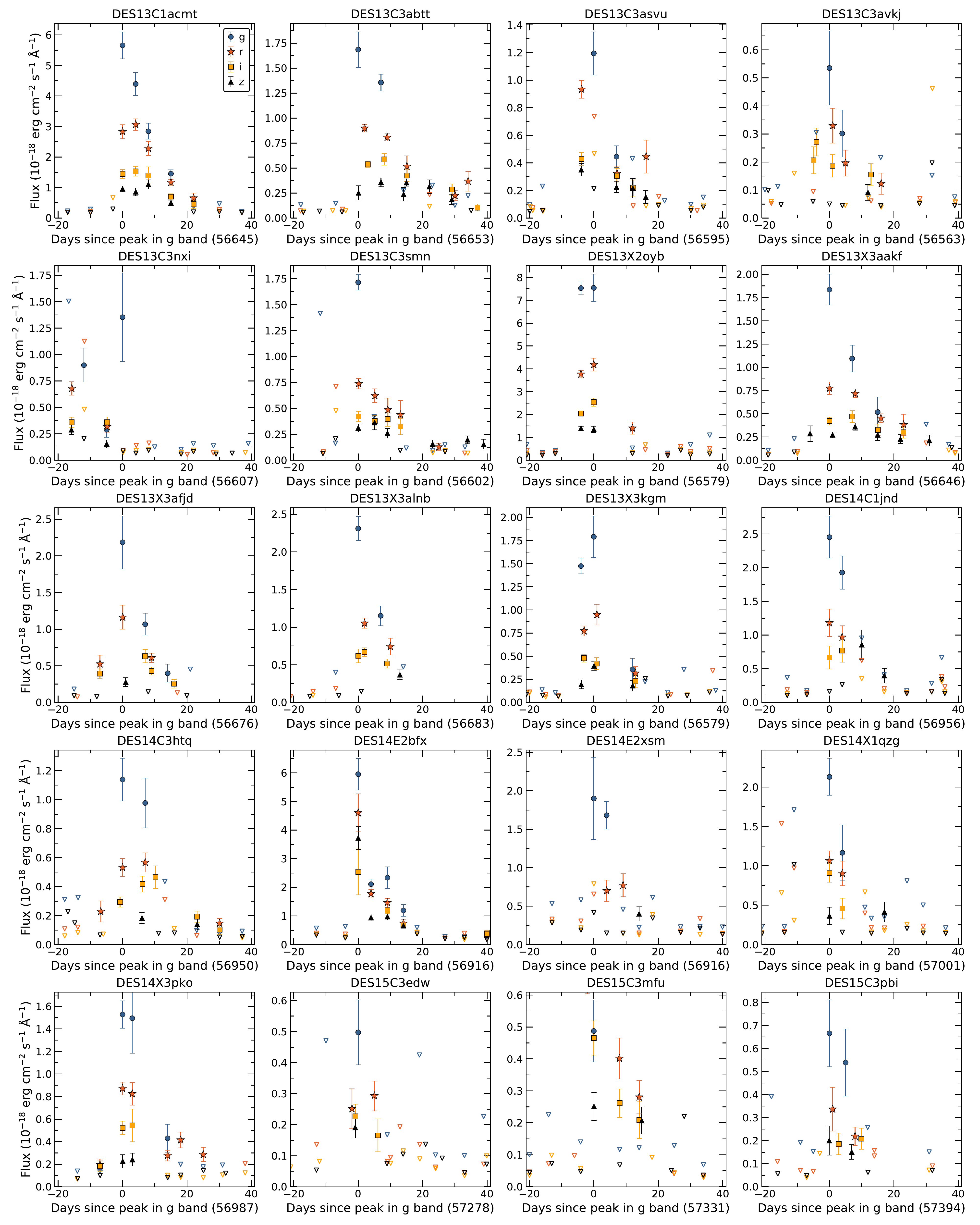}
\captionsetup{labelfont=bf, font=small, justification=centering, labelsep=period}
\captionof{figure}{Observer frame light curves of bronze sample transients. Open triangles are the same as in Figure \ref{fig:gold_lcs_all}.}
\label{fig:bronze_lcs_all}
\end{minipage}
\clearpage
\addtocounter{figure}{-1}

\begin{minipage}{\textwidth}
\includegraphics[width=0.98\textwidth]{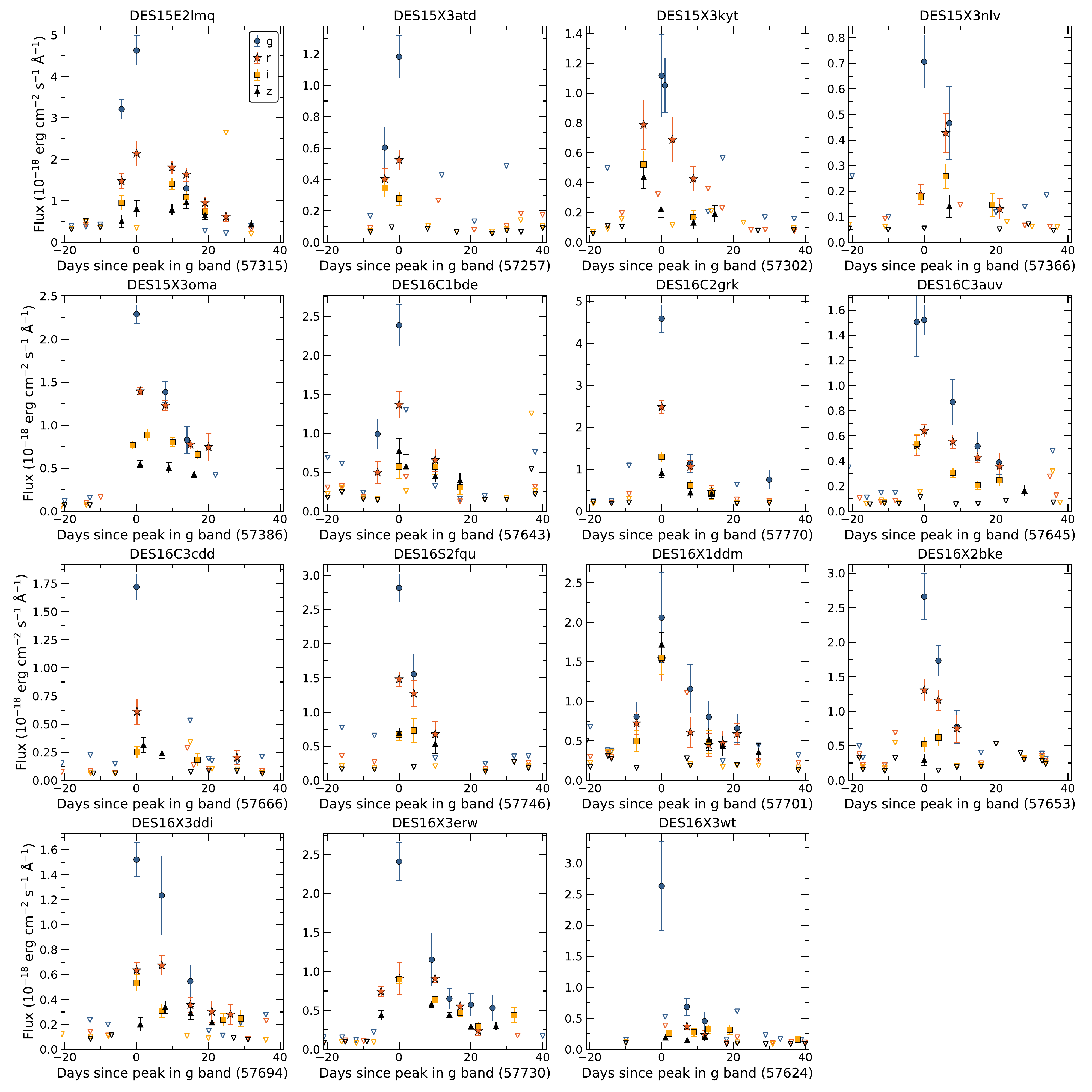}
\captionsetup{labelfont=bf, font=small, justification=centering, labelsep=period}
\captionof{figure}{(Continued.)}
\label{fig:bronze_lcs_all2}
\end{minipage}


\bsp	
\label{lastpage}
\end{document}